\documentclass[pdflatex, sn-nature, iicol]{sn-jnl}% Style for submissions to Nature Portfolio journals
\usepackage{graphicx}%
\usepackage{multirow}%
\usepackage{amsmath,amssymb,amsfonts,amsthm,mathrsfs}

\usepackage[fleqn]{nccmath}
\usepackage[title]{appendix}%
\usepackage{xcolor}%
\usepackage{textcomp}%
\usepackage{manyfoot}%
\usepackage{booktabs}%
\usepackage{algorithm}%
\usepackage{algorithmicx}%
\usepackage{algpseudocode}%
\usepackage{listings}%

\usepackage[english]{babel}
\usepackage[utf8]{inputenc}
\usepackage[T1]{fontenc}
\usepackage[normalem]{ulem} %for strikethrough

\usepackage[left]{lineno}
\usepackage{graphicx}
\graphicspath{ {./Figures/}}

\usepackage{caption}
\usepackage{subcaption}

\usepackage{nicematrix, booktabs, varwidth}
\usepackage{siunitx}
\DeclareSIUnit\we{w.e.}
\DeclareSIUnit\lightspeed{c}
\DeclareSIUnit\omflux{cm^{-2} s^{-1}}
\DeclareSIUnit\Bqconc{Bq \, m^{-3}}
\DeclareSIUnit{\keVee}{keV\textsubscript{ee}}

\usepackage[version=4]{mhchem}
\usepackage{csquotes}
\usepackage{url}
\usepackage{multirow}
\usepackage[export]{adjustbox}
\usepackage{orcidlink}

\newcommand{\nucleus}{NUCLEUS}
\newcommand{\Geant}{\textsc{Geant4}}
\newcommand{\cenns}{CE$\nu$NS}
\newcommand{\alo}{\ce{Al2O3}}
\newcommand{\cawo}{\ce{CaWO4}}
\newcommand{\bfourc}{\ce{B4C}}
\newcommand{\rate}[1]{\SI{#1}{\per\day\per\kilo\gram\per\kilo\electronvolt}}

\usepackage{footmisc} % allows footnote symbols

\usepackage[final]{changes}

\theoremstyle{thmstyleone}%
\theoremstyle{thmstyletwo}%

\theoremstyle{thmstylethree}%

\begin{document}

\title[Particle background characterization and prediction for the NUCLEUS reactor CEvNS experiment]{Particle background characterization and prediction for the \nucleus{} reactor \cenns{} experiment}

% Author list Version_8
%======================

% BEGINNING OF COPY/PASTE FROM THE OFFICIAL LIST OF AUTHORS

% Author list

% to choose the style of affiliation, un-comment one of 
% the \instByName values below 

\newcount\instByName

% \instByName=1 % institutes by names as superscript for author
\instByName=0 % institutes by numbers as superscript for author

% definitions of institutes
%==========================

% \affil[1]{\orgdiv{Department}, \orgname{Organization}, \orgaddress{\street{Street}, \city{City}, \postcode{100190}, \state{State}, \country{Country}}}

% \newcommand{}{
%     \orgdiv{}, 
%     \orgname{}, 
%     \orgaddress{%
%         \street{}, 
%         \city{}, 
%         \postcode{}, 
%         \state{}, 
%         \country{}
%         }
%     }

% Austria

\newcommand{\iHEPHY}{%
    % \orgdiv{}, 
    \orgname{Institut f\"ur Hochenergiephysik der \"Osterreichischen Akademie der Wissenschaften}, 
    \orgaddress{%
        \street{Dominikanerbastei~16}, 
        \city{Wien}, 
        \postcode{A-1010}, 
        % \state{}, 
        \country{Austria}%
        }%
    }

\newcommand{\iTUW}{%
    \orgdiv{Atominstitut}, 
    \orgname{Technische Universit\"at Wien}, 
    \orgaddress{%
        \street{Stadionallee~2}, 
        \city{Wien}, 
        \postcode{A-1020}, 
        % \state{}, 
        \country{Austria}%
        }%
    }

% France

\newcommand{\iCEA}{%
    \orgdiv{IRFU}, 
    \orgname{CEA, Universit\'{e} Paris-Saclay}, 
    \orgaddress{%
        \street{B\^{a}timent 141}, 
        \city{Gif-sur-Yvette}, 
        \postcode{F-91191}, 
        % \state{}, 
        \country{France}%
        }%
    }

\newcommand{\iEdF}{%
    \orgdiv{Centre nucl\'eaire de production d'\'electricit\'e de Chooz, Service Automatismes-Essais}, 
    \orgname{\'Electricit\'e de France}, 
    \orgaddress{%
        \street{}, 
        \city{Givet}, 
        \postcode{F-08600}, 
        % \state{}, 
        \country{France}%
        }%
    }

% Germany

\newcommand{\iMPIK}{%
    % \orgdiv{}, 
    \orgname{Max-Planck-Institut für Kernphysik}, 
    \orgaddress{%
        \street{Saupfercheckweg 1}, 
        \city{Heidelberg}, 
        \postcode{D-69117}, 
        % \state{}, 
        \country{Germany}%
        }%
    }

\newcommand{\iMPP}{%
    % \orgdiv{}, 
    \orgname{Max-Planck-Institut f\"ur Physik}, 
    \orgaddress{%
        \street{Boltzmannstra{\ss}e~8}, 
        \city{Garching}, 
        \postcode{D-85748}, 
        \country{Germany}%
        }%
    }

\newcommand{\iTUM}{%
    \orgdiv{Physik-Department}, 
    \orgname{Technische Universit\"at M\"unchen}, 
    \orgaddress{%
        \street{James-Franck-Straße 1}, 
        \city{Garching}, 
        \postcode{D-85748}, 
        % \state{}, 
        \country{Germany}%
        }%
    }

% Italy

\newcommand{\iINFNRoma}{%
    % \orgdiv{}, 
    \orgname{Istituto Nazionale di Fisica Nucleare -- Sezione di Roma}, 
    \orgaddress{%
        \street{Piazzale Aldo Moro 2}, 
        \city{Roma}, 
        \postcode{I-00185}, 
        % \state{}, 
        \country{Italy}%
        }%
    }

\newcommand{\iSapienza}{%
    \orgdiv{Dipartimento di Fisica}, 
    \orgname{Sapienza Universit\`{a} di Roma}, 
    \orgaddress{%
        \street{Piazzale Aldo Moro 5}, 
        \city{Roma}, 
        \postcode{I-00185}, 
        % \state{}, 
        \country{Italy}%
        }%
    }
    
\newcommand{\iINFNTorVergata}{%
    % \orgdiv{}, 
    \orgname{Istituto Nazionale di Fisica Nucleare -- Sezione di Roma "Tor Vergata"}, 
    \orgaddress{%
        \street{Via della Ricerca Scientifica 1}, 
        \city{Roma}, 
        \postcode{I-00133}, 
        % \state{}, 
        \country{Italy}%
        }%
    }

\newcommand{\iTorVergata}{%
    \orgdiv{Dipartimento di Fisica}, 
    \orgname{Universit\`{a} di Roma "Tor Vergata"}, 
    \orgaddress{%
        \street{Via della Ricerca Scientifica 1}, 
        \city{Roma}, 
        \postcode{I-00133}, 
        % \state{}, 
        \country{Italy}%
        }%
    }

\newcommand{\iCNR}{%
    \orgdiv{Istituto di Nanotecnologia}, 
    \orgname{Consiglio Nazionale delle Ricerche}, 
    \orgaddress{%
        \street{Piazzale Aldo Moro 5}, 
        \city{Roma}, 
        \postcode{I-00185}, 
        % \state{}, 
        \country{Italy}%
        }%
    }

\newcommand{\iINFNFerrara}{%
    % \orgdiv{}, 
    \orgname{Istituto Nazionale di Fisica Nucleare -- Sezione di Ferrara}, 
    \orgaddress{%
        \street{Via Giuseppe Saragat 1c}, 
        \city{Ferrara}, 
        \postcode{I-44122}, 
        % \state{}, 
        \country{Italy}%
        }%
    }

\newcommand{\iFerrara}{%
    \orgdiv{Dipartimento di Fisica}, 
    \orgname{Universit\`{a} di Ferrara}, 
    \orgaddress{%
        \street{Via Giuseppe Saragat 1}, 
        \city{Ferrara}, 
        \postcode{I-44122}, 
        % \state{}, 
        \country{Italy}%
        }%
    }

\newcommand{\iINFNLnGS}{%
    % \orgdiv{}, 
    \orgname{Istituto Nazionale di Fisica Nucleare -- Laboratori Nazionali del Gran Sasso}, 
    \orgaddress{%
        \street{Via Giovanni Acitelli 22}, 
        \city{Assergi (L’Aquila)}, 
        \postcode{I-67100}, 
        % \state{}, 
        \country{Italy}%
        }%
    }

\newcommand{\iBicocca}{%
    \orgdiv{Dipartimento di Fisica}, 
    \orgname{Universit\`{a} di Milano Bicocca}, 
    \orgaddress{%
        \street{}, 
        \city{Milan}, 
        \postcode{I-20126}, 
        % \state{}, 
        \country{Italy}%
        }%
    }

% Portugal

\newcommand{\iCoimbra}{%
    \orgdiv{LIBPhys-UC, Departamento de Fisica}, 
    \orgname{Universidade de Coimbra}, 
    \orgaddress{%
        \street{Rua Larga 3004-516}, 
        \city{Coimbra}, 
        \postcode{P-3004-516}, 
        % \state{}, 
        \country{Portugal}%
        }%
    }

% the style of affiliation superscript
\ifnum\instByName=1
    % institutes by names as superscript for author
    
    % Austria
    \newcommand{\HEPHY}{HEPHY}
    \newcommand{\TUW}{TUW}

    % France
    \newcommand{\CEA}{CEA}
    \newcommand{\EdF}{EdF}

    %Germany
    \newcommand{\MPIK}{MPIK}
    \newcommand{\MPP}{MPP}
    \newcommand{\TUM}{TUM}

    % Italy
    \newcommand{\INFNRoma}{INFNRoma}
    \newcommand{\Sapienza}{Sapienza}
    \newcommand{\INFNTorVergata}{INFNTorVergata}
    \newcommand{\TorVergata}{TorVergata}
    \newcommand{\CNR}{CNR}
    \newcommand{\INFNFerrara}{INFNFerrara}
    \newcommand{\Ferrara}{Ferrara}
    \newcommand{\INFNLnGS}{INFNLnGS}
    \newcommand{\Bicocca}{Bicocca}    

    % Portugal
    \newcommand{\Coimbra}{Coimbra}
    
\else    
    % institutes by numbers as superscript for author
    % numbers have to be set correctly, 
    % according to the institutes of authors in the alphabetical order
    % the affil (see affil commands) should be in the increasing index
    
    % Austria
    \newcommand{\HEPHY}{3}
    \newcommand{\TUW}{1}

    % France
    \newcommand{\CEA}{5}
    \newcommand{\EdF}{ERROR}

    %Germany
    \newcommand{\MPIK}{ERROR}
    \newcommand{\MPP}{2}
    \newcommand{\TUM}{8}

    % Italy
    \newcommand{\INFNRoma}{6}
    \newcommand{\Sapienza}{7}
    
    \newcommand{\INFNTorVergata}{4}
    \newcommand{\TorVergata}{9}
    
    \newcommand{\CNR}{ERROR}
    
    \newcommand{\INFNFerrara}{10}
    \newcommand{\Ferrara}{ERROR}
    
    \newcommand{\INFNLnGS}{ERROR}
    
    \newcommand{\Bicocca}{ERROR}    

    % Portugal
    \newcommand{\Coimbra}{ERROR}

    %Now at
    \newcommand{\alsoatCoimbra}{\dag}
    \newcommand{\nowatMPIK}{\ddag}
    \newcommand{\nowatMPP}{$\mathsection$}

\fi

% definitions of authors
%=======================


% \author[]{
%     \fnm{} 
%     \sur{} 
%     \email{\newline , : }  
%     \orcidlink{}
%     }

% \author[\TUW]{H.~Abele~\orcidlink{0000-0002-6832-9051}}
\author[\TUW]{%
    \fnm{H.} 
    \sur{Abele} 
    %\email{\newline Abele, H.: hartmut.abele@tuwien.ac.at}  
    \orcidlink{0000-0002-6832-9051}%
    }

% \author[\MPP]{G.~Angloher}
\author[\MPP]{%
    \fnm{G.} 
    \sur{Angloher} 
     %\email{\newline Angloher, G.: gangloher@epo.org}
    }

% \author[\HEPHY]{B.~Arnold}
\author[\HEPHY]{%
    \fnm{B.}
    \sur{Arnold}
    %\email{\newline Arnold B.: bernhard.arnold@oeaw.ac.at}
    }
    
% \author[\INFNTorVergata]{M.~Atzori~Corona~\orcidlink{0000-0001-5092-3602}}
\author[\INFNTorVergata]{%
    \fnm{M.}
    \sur{Atzori~Corona}
    %\email{\newline Atzori~Corona, M.: mcorona@roma2.infn.it}
    \orcidlink{0000-0001-5092-3602}%
    }

% \author[\MPP, \thanks{Also at: \iCoimbra}]{A.~Bento~\orcidlink{0000-0002-3817-6015}}
\author[\MPP]{%
    \fnm{A.}
    \sur{Bento}
    %\email{\newline Bento A.: bento@mpp.mpg.de}
    \orcidlink{0000-0002-3817-6015}
    \textsuperscript{\alsoatCoimbra,\,}%
    }

% \author[\CEA]{E.~Bossio~\orcidlink{0000-0001-9304-1829}}
\author[\CEA]{%
    \fnm{E.}
    \sur{Bossio}
    %\email{\newline Bossio E.: elisabetta.bossio@cea.fr}
    \orcidlink{0000-0001-9304-1829}%
    }
    
% \author[\HEPHY]{F.~Buchsteiner}
\author[\HEPHY]{%
    \fnm{F.}
    \sur{Buchsteiner}%
    %\email{\newline Buchsteiner, F. florian.buchsteiner@oeaw.ac.at}
    }
    
% \author[\HEPHY]{J.~Burkhart~\orcidlink{0000-0002-1989-7845}}
\author[\HEPHY]{%
    \fnm{J.}
    \sur{Burkhart}
    %\email{\newline Burkhart, J.: jens.burkhart@cern.ch}
    \orcidlink{0000-0002-1989-7845}%
    }
    
% \author[\MPP, \thanks{Now at: \iBicocca}]{L.~Canonica~\orcidlink{0000-0001-8734-206X}} %V7

% \author[\INFNRoma]{F.~Cappella~\orcidlink{0000-0003-0900-6794}}
\author[\INFNRoma]{%
    \fnm{F.}
    \sur{Cappella}
    %\email{\newline Cappella, F.: fabio.cappella@roma1.infn.it}
    \orcidlink{0000-0003-0900-6794}%
    }
    
% \author[\Sapienza, \INFNRoma]{M.~Cappelli~\orcidlink{0009-0002-6148-5964}}
\author[\Sapienza, \INFNRoma]{%
    \fnm{M.}
    \sur{Cappelli}
    %\email{\newline Cappelli, M.: matteo.cappelli@roma1.infn.it}
    \orcidlink{0009-0002-6148-5964}%
    }
    
% \author[\INFNRoma]{L.~Cardani} %V5

% \author[\INFNRoma]{N.~Casali~\orcidlink{0000-0003-3669-8247}}
\author[\INFNRoma]{%
    \fnm{N.}
    \sur{Casali}
    %\email{\newline Casali, N.: nicola.casali@roma1.infn.it}
    \orcidlink{0000-0003-3669-8247}%
    }
    
% \author[\INFNTorVergata]{R.~Cerulli~\orcidlink{0000-0003-2051-3471}}
\author[\INFNTorVergata]{%
    \fnm{R.}
    \sur{Cerulli}
    %\email{\newline Cerulli, R.: riccardo.cerulli@roma2.infn.it}
    \orcidlink{0000-0003-2051-3471}%
    }
    
% \author[\CNR, \INFNRoma]{I.~Colantoni} %V5

% \author[\INFNRoma]{A.~Cruciani~\orcidlink{0000-0003-2247-8067}}
\author[\INFNRoma]{%
    \fnm{A.}
    \sur{Cruciani}
    %\email{\newline Cruciani, A.: angelo.cruciani@roma1.infn.it}
    \orcidlink{0000-0003-2247-8067}%
    }
    
% \author[\INFNRoma]{G.~Del~Castello~\orcidlink{0000-0001-7182-358X}}
\author[\INFNRoma]{%
    \fnm{G.}
    \sur{Del~Castello}
    %\email{\newline Del~Castello, G.: Giorgio.DelCastello@roma1.infn.it}
    \orcidlink{0000-0001-7182-358X}%
    }

% \author[\Sapienza, \INFNRoma]{M.~del~Gallo~Roccagiovine}
\author[\Sapienza, \INFNRoma]{%
    \fnm{M.}
    \sur{del~Gallo~Roccagiovine}%
    %\email{\newline del~Gallo~Roccagiovine, M.: matteo.delgalloroccagiovine@roma1.infn.it}
    }
    
% \author[\TUW]{A.~Doblhammer} %V7

% \author[\TUW]{S.~Dorer~\orcidlink{0009-0001-1670-5780}}
\author[\TUW]{%
    \fnm{S.}
    \sur{Dorer}
    %\email{\newline Dorer, S.: sebastian.dorer@tuwien.ac.at}
    \orcidlink{0009-0001-1670-5780}%
    }
    
% \author[\TUM]{A.~Erhart~\orcidlink{0000-0002-8721-177X}}
\author[\TUM]{%
    \fnm{A.}
    \sur{Erhart}
    %\email{\newline Erhart, A.: andreas.erhart@tum.de}
    \orcidlink{0000-0002-8721-177X}%
    }
    
% \author[\HEPHY]{M.~Friedl~\orcidlink{0000-0002-7420-2559}}
\author[\HEPHY]{%
    \fnm{M.}
    \sur{Friedl}
    %\email{\newline Friedl, M.: markus.friedl@oeaw.ac.at}
    \orcidlink{0000-0002-7420-2559}}%

% \author[\HEPHY]{S.~Fichtinger}
\author[\HEPHY]{%
    \fnm{S.}
    \sur{Fichtinger}%
    %\email{\newline Fichtinger, S.: stephan.fichtinger@oeaw.ac.at}
    }

% \author[\MPP]{A.~Garai} %V7

% \author[\HEPHY]{V.M.~Ghete~\orcidlink{0000-0002-9595-6560}}
\author[\HEPHY]{%
    \fnm{V.M.}
    \sur{Ghete}
    %\email{\newline Ghete, V. M.: Vasile-Mihai.Ghete@oeaw.ac.at}
    \orcidlink{0000-0002-9595-6560}%
    }

% \author[\TorVergata, INFNTorVergata]{M.~Giammei~\orcidlink{0009-0006-9104-2055}}
\author[\TorVergata, \INFNTorVergata]{%
    \fnm{M.}
    \sur{Giammei}
    %\email{\newline Giammei, M: giammeim@roma2.infn.it}
    \orcidlink{0009-0006-9104-2055}%
    }

% \author[\CEA, \thanks{Now at: \MPIK}]{C.~Goupy~\orcidlink{0000-0003-4954-5311}}
\author*[\CEA]{%
    \fnm{C.} 
    \sur{Goupy}
    % \email{chloe.goupy@mpi-hd.mpg.de}  
    \orcidlink{0000-0003-4954-5311}
    \textsuperscript{\nowatMPIK,\,}%
    }\email{chloe.goupy@mpi-hd.mpg}

%\author[\Ferrara, \INFNFerrara]{V.~Guidi} %V5

% \author[\MPP, \TUM]{D.~Hauff}
\author[\MPP, \TUM]{%
    \fnm{D.}
    \sur{Hauff}%
    %\email{\newline Hauff, D.: hauff@mpp.mpg.de}
    }
    
% \author[\CEA]{F.~Jeanneau~\orcidlink{0000-0002-6360-6136}}
\author[\CEA]{%
    \fnm{F.}
    \sur{Jeanneau}
    %\email{\newline Jeanneau, F.: fabien.jeanneau@cea.fr}
    \orcidlink{0000-0002-6360-6136}%
    }

% \author[\TUW]{E.~Jericha~\orcidlink{0000-0002-8663-0526}}
\author[\TUW]{%
    \fnm{E.}
    \sur{Jericha}%
    %\email{\newline Jericha, E.: erwin.jericha@tuwien.ac.at}~\orcidlink{0000-0002-8663-0526}
    }

% \author[\TUM]{M.~Kaznacheeva~\orcidlink{0000-0002-2712-1326}}
\author[\TUM]{%
    \fnm{M.}
    \sur{Kaznacheeva}
    %\email{\newline Kaznacheeva, M.: margarita.kaznacheeva@tum.de}
    \orcidlink{0000-0002-2712-1326}%
    }

% \author[\TUM]{A.~Kinast~\orcidlink{0000-0002-5894-2303}} %V8

% \author[\HEPHY]{H.~Kluck~\orcidlink{0000-0003-3061-3732}}
\author[\HEPHY]{%
    \fnm{H.} 
    \sur{Kluck} 
    %\email{\newline Kluck, H.: Holger.Kluck@oeaw.ac.at}  
    \orcidlink{0000-0003-3061-3732}%
    }

% \author[\MPP]{A.~Langenk\"{a}mper}
\author[\MPP]{%
    \fnm{A.} 
    \sur{Langenk\"{a}mper} 
    %\email{\newline Langenk\"{a}mper, A.: langenk@mpp.mpg.de}  
    }

% \author[\CEA, \TUM, \thanks{Now also at: \MPIK}]{T.~Lasserre~\orcidlink{0000-0002-4975-2321}}
\author[\CEA, \TUM]{%
    \fnm{T.} 
    \sur{Lasserre}
    %\email{\newline Lasserre, T.: thierry.lasserre@mpi-hd.mpg.de}  
    \orcidlink{0000-0002-4975-2321}
    \textsuperscript{\nowatMPIK,\,}%
    }

% \author[\CEA]{D.~Lhuillier~\orcidlink{0000-0003-2324-0149}}
\author[\CEA]{%
    \fnm{D.} 
    \sur{Lhuillier} 
    %\email{\newline Lhuillier, D.: david.lhuillier@cea.fr}  
    \orcidlink{0000-0003-2324-0149}
    }

% \author[\MPP]{M.~Mancuso~\orcidlink{0000-0001-9805-475X}}
\author[\MPP]{%
    \fnm{M.} 
    \sur{Mancuso} 
    %\email{\newline Mancuso, M.: michele.mancuso@mpp.mpg.de}  
    \orcidlink{0000-0001-9805-475X}
    }

% \author[\CEA, \TUW]{R.~Martin}
\author[\CEA, \TUW]{%
    \fnm{R.} 
    \sur{Martin} 
    %\email{\newline Martin,R.: romain.martin2@cea.fr}  
    % \orcidlink{}
    }

% \author[\MPP]{B.~Mauri}
\author[\MPP]{%
    \fnm{B.} 
    \sur{Mauri} 
    %\email{\newline Mauri, B.: bmauri@mpp.mpg.de}  
    % \orcidlink{}
    }

% \author[\INFNFerrara]{A.~Mazzolari}
\author[\INFNFerrara]{%
    \fnm{A.} 
    \sur{Mazzolari} 
    %\email{\newline Mazzolari, A.: andrea.mazzolari@cern.ch}  
    % \orcidlink{}
    }

% \author[\CEA]{L.~McCallin}
\author[\CEA]{%
    \fnm{L.} 
    \sur{McCallin} 
    %\email{\newline McCallin, L.: liliane.mccallin@cea.fr}  
    % \orcidlink{}
    }

% \author[\CEA]{E.~Mazzucato} %V7

% \author[\CEA]{H.~Neyrial}
\author[\CEA]{%
    \fnm{H.} 
    \sur{Neyrial} 
    %\email{\newline Neyrial, H.: Hubert.NEYRIAL@cea.fr}  
    % \orcidlink{}
    }

% \author[\CEA]{C.~Nones}
\author[\CEA]{%
    \fnm{C.} 
    \sur{Nones} 
    %\email{\newline Nones, C.: Claudia.Nones@cea.fr}  
    % \orcidlink{}
    }

% \author[\TUM]{L.~Oberauer}
\author[\TUM]{%
    \fnm{L.} 
    \sur{Oberauer} 
    %\email{\newline Oberauer, L.: lothar.oberauer@tum.de}  
    % \orcidlink{}
    }

% \author[\TUM]{T.~Ortmann, \thanks{Now at: \MPP}}
\author[\TUM]{%
    \fnm{T.} 
    \sur{Ortmann}
    \textsuperscript{\nowatMPP,\,}
    %\email{\newline Ortmann, T.: tobias.ortmann@tum.de}  
    % \orcidlink{}
    }

% \author[\INFNLnGS, \thanks{Now at: \iBicocca}]{L.~Pattavina~\orcidlink{0000-0003-4192-849X}} %V8

% \author[\TUM, \CEA]{L.~Peters~, \thanks{Now at: \iMPIK}\orcidlink{0000-0002-1649-8582}}
\author[\TUM, \CEA]{%
    \fnm{L.} 
    \sur{Peters} 
    %\email{\newline Peters, L.: lilly.peters@tum.de}  
    \orcidlink{0000-0002-1649-8582}
    \textsuperscript{\nowatMPIK,\,}
    }

% \author[\MPP]{F.~Petricca~\orcidlink{0000-0002-6355-2545}}
\author[\MPP]{%
    \fnm{F.} 
    \sur{Petricca} 
    %\email{\newline Petricca, F.: petricca@mpp.mpg.de}  
    \orcidlink{0000-0002-6355-2545}
    }

% \author[\TUM]{W.~Potzel}
\author[\TUM]{%
    \fnm{W.} 
    \sur{Potzel} 
    %\email{\newline Potzel, W.: walter.potzel@tum.de}  
    % \orcidlink{}
    }

% \author[\MPP]{F.~Pr\"{o}bst}
\author[\MPP]{%
    \fnm{F.} 
    \sur{Pr\"{o}bst} 
    %\email{\newline Pr\"{o}bst, F.: proebst@mpp.mpg.de}  
    % \orcidlink{}
    }

% \author[\MPP]{F.~Pucci}
\author[\MPP]{%
    \fnm{F.} 
    \sur{Pucci} 
    %\email{\newline Pucci, F.: francesca.pucci@lngs.infn.it}  
    % \orcidlink{}
    }

% \author[\HEPHY, TUW]{F.~Reindl~\orcidlink{0000-0003-0151-2174}}
\author[\HEPHY, \TUW]{%
    \fnm{F.} 
    \sur{Reindl} 
    %\email{\newline Reindl, F.: florian.reindl@tuwien.ac.at}  
    \orcidlink{0000-0003-0151-2174}
    }

% \author[\CEA]{R.~Rogly} %V5

% \author[\INFNFerrara]{M.~Romagnoni}
\author[\INFNFerrara]{%
    \fnm{M.} 
    \sur{Romagnoni} 
    %\email{\newline Romagnoni, M.: romagnoni@fe.infn.it}  
    % \orcidlink{}
    }

% \author[\TUM]{J.~Rothe~\orcidlink{0000-0001-5748-7428}}
\author[\TUM]{%
    \fnm{J.} 
    \sur{Rothe} 
    %\email{\newline Rothe, J.: johannes.rothe@tum.de}  
    \orcidlink{0000-0001-5748-7428}
    }

% \author[\TUM]{N.~Schermer~\orcidlink{0009-0004-4213-5154}}
\author[\TUM]{%
    \fnm{N.} 
    \sur{Schermer} 
    %\email{\newline Schermer, N.: nicole.schermer@tum.de}  
    \orcidlink{0009-0004-4213-5154}
    }

% \author[\HEPHY, \TUW]{J.~Schieck~\orcidlink{0000-0002-1058-8093}}
\author[\HEPHY, \TUW]{%
    \fnm{J.} 
    \sur{Schieck} 
    %\email{\newline Schieck, J.: Jochen.Schieck@oeaw.ac.at}  
    \orcidlink{0000-0002-1058-8093}
    }

% \author[\TUM]{S.~Sch\"{o}nert~\orcidlink{0000-0001-5276-2881}}
\author[\TUM]{%
    \fnm{S.} 
    \sur{Sch\"{o}nert} 
    %\email{\newline Sch\"{o}nert, S.: schoenert@ph.tum.de}  
    \orcidlink{0000-0001-5276-2881}
    }

% \author[\HEPHY, TUW]{C.~Schwertner}
\author[\HEPHY, \TUW]{%
    \fnm{C.} 
    \sur{Schwertner} 
    %\email{\newline Schwertner, C.: Christoph.Schwertner@oeaw.ac.at}  
    % \orcidlink{}
    }

% \author[\CEA]{L.~Scola}
\author[\CEA]{%
    \fnm{L.} 
    \sur{Scola} 
    %\email{\newline Scola, L.: loris.scola@cea.fr}  
    % \orcidlink{}
    }

% \author[\CEA]{G.~Soum-Sidikov~\orcidlink{0000-0003-1900-1794}}
\author[\CEA]{%
    \fnm{G.} 
    \sur{Soum-Sidikov} 
    %\email{\newline Soum-Sidikov, G.: gabrielle.soum@cea.fr}  
    \orcidlink{0000-0003-1900-1794}
    }

% \author[\MPP]{L.~Stodolsky}
\author[\MPP]{%
    \fnm{L.} 
    \sur{Stodolsky} 
    %\email{\newline Stodolsky, L.: les@mpp.mpg.de}  
    % \orcidlink{}
    }

% \author[\TUM]{R.~Strauss~\orcidlink{0000-0002-5589-9952}}
\author[\TUM]{%
    \fnm{R.} 
    \sur{Strauss} 
    %\email{\newline Strauss, R.: raimund.strauss@tum.de}  
    \orcidlink{0000-0002-5589-9952}
    }

% \author[\Ferrara, \INFNFerrara]{M.~Tamisari} %V8

% \author[\HEPHY]{R.~Thalmeier~\orcidlink{0009-0003-4480-0990}}
\author[\HEPHY]{%
    \fnm{R.} 
    \sur{Thalmeier} 
    %\email{\newline Thalmeier, R.: richard.thalmeier@oeaw.ac.at}  
    \orcidlink{0009-0003-4480-0990}
    }

% \author[\INFNRoma]{C.~Tomei}
\author[\INFNRoma]{%
    \fnm{C.} 
    \sur{Tomei} 
    %\email{\newline Tomei, C.: claudia.tomei@roma1.infn.it}  
    % \orcidlink{}
    }

% \author[\Sapienza, \INFNRoma]{M.~Vignati~\orcidlink{0000-0002-8945-1128}}
\author[\Sapienza, \INFNRoma]{%
    \fnm{M.} 
    \sur{Vignati} 
    %\email{\newline Vignati, M.: Marco.Vignati@roma1.infn.it}  
    \orcidlink{0000-0002-8945-1128}
    }

% \author[\CEA]{M.~Vivier~\orcidlink{0000-0003-2199-0958}}
\author[\CEA]{%
    \fnm{M.} 
    \sur{Vivier} 
    %\email{\newline Vivier, M.: Matthieu.Vivier@cea.fr}  
    \orcidlink{0000-0003-2199-0958}
    }

% \author[\TUM]{V.~Wagner~\orcidlink{0000-0003-1845-4951}}

\author[\TUM]{%
    \fnm{A.}
    \sur{Wex}
    %\email{\newline Wex A.: alexander.wex@tum.de}
    \orcidlink{0009-0003-5371-2466}
}

% force new page - comment it when used in papers

    \affil[\TUW]{\iTUW}
    \affil[\MPP]{\iMPP}
    \affil[\HEPHY]{\iHEPHY}
    \affil[\INFNTorVergata]{\iINFNTorVergata}
    \affil[\CEA]{\iCEA}
    \affil[\INFNRoma]{\iINFNRoma}
    \affil[\Sapienza]{\iSapienza}
    \affil[\TUM]{\iTUM}
    \affil[\TorVergata]{\iTorVergata}
    \affil[\INFNFerrara]{\iINFNFerrara}

    % no authors affiliated to:
    
    % \affil[\EdF]{\iEdF}
    % \affil[\MPIK]{\iMPIK}
    % \affil[\CNR]{\iCNR}
    % \affil[\Ferrara]{\iFerrara}
    % \affil[\INFNLnGS]{\iINFNLnGS}
    % \affil[\Bicocca]{\iBicocca}    
    % \affil[\Coimbra]{\iCoimbra}

    %Now at affiliations
    \affil[\alsoatCoimbra]{\small Also at \iCoimbra}
    \affil[\nowatMPP]{\small Now at \iMPP}
    \affil[\nowatMPIK]{\small Now at \iMPIK}
    % \footnotetext[4]{Now at Dipartimento di Fisica, Universit\`{a} di Milano Bicocca, I-20126, Milan, Italy}

% END OF COPY/PASTE FROM THE OFFICIAL LIST OF AUTHORS


\abstract{\nucleus{} is a cryogenic detection experiment which aims to measure \textit{Coherent Elastic Neutrino–Nucleus Scattering} (\cenns{}) and to search for new physics at the Chooz nuclear power plant in France. 
This article reports on the prediction of particle-induced backgrounds, especially focusing on the sub-keV energy range, which is a poorly known region where most of the \cenns{} signal from reactor antineutrinos is expected. 
Together with measurements of the environmental background radiations at the experimental site, extensive Monte Carlo simulations based on the \Geant{} package were run both to optimize the experimental setup for background reduction and to estimate the residual rates arising from different contributions such as cosmic ray-induced radiations, environmental gammas and material radioactivity. 
The \nucleus{} experimental setup is predicted to achieve a total rejection power of more than two orders of magnitude, leaving a residual background component which is strongly dominated by cosmic ray-induced neutrons.
In the \cenns{} signal region of interest between 10 and 100\,eV, a total particle background rate of $\sim$\,\rate{250} is expected in the \cawo{} target detectors. 
This corresponds to a signal-to-background ratio $\gtrsim$ 1, and therefore meets the required specifications in terms of particle background rejection for the detection of reactor antineutrinos through \cenns{}.
}

\keywords{Coherent Elastic Neutrino-Nucleus Scattering (\cenns{}), neutrino-nucleus interactions, nuclear reactor, cosmic-ray induced background, gamma ray spectroscopy, material contamination measurement, low background, shielding strategy, sub-keV energies}



\maketitle

%line numbering 
%\linenumbers
%\modulolinenumbers[2]

\section{Introduction}\label{sec:intro}
Coherent Elastic Neutrino-Nucleus Scattering (\cenns{}) is a neutral current process predicted in 1974~\cite{freedman1974coherent,kopelievichfrankfurt1974coherent} soon after the first observation of weak neutral currents by the Gargamelle experiment \cite{Gargamelle1973}. 
This process is effective at low momentum transfers, allowing neutrinos of any flavor with energies up to $\sim$\,50\,MeV to coherently scatter over a target nucleus as a whole, giving rise to a significant enhancement to the cross-section~\cite{drukier1984principles}. 
This coherent enhancement is proportional to the squared number of neutrons in the target nucleus, which favors heavy target nuclei for studying the process. 
It opens up the possibility of using smaller detector payloads for the detection of neutrinos as well as for high-precision searches of physics beyond the Standard Model. 
However, the detection of \cenns{} poses multiple challenges, as the experimental signature is a standalone nuclear recoil. \deleted{Even in the most favorable cases, such} For pion-at-rest 30\,MeV neutrinos scattering on sodium, the recoil energies reach only up to $\sim$\,100\,keV, while \deleted{typical} reactor neutrinos of \deleted{about} \added{maximum} 10\,MeV interacting with heavier targets such as germanium produce recoils \deleted{of merely} \added{not exceeding} $\sim$\,3 keV.

The process would then remain elusive for more than 40 years until the COHERENT collaboration observed it for the first time on Cs and I target nuclei, using the Spallation Neutron Source (SNS) facility (Oak Ridge, Tennessee) as a pulsed source of neutrinos with an average energy of $\mathrm{\sim}$30\,MeV~\cite{akimov2017observation, akimov2022measurement}.
Detection on Ar~\cite{akimov2021first} and Ge~\cite{adamski2024first} nuclei later followed at the SNS from the same collaboration and confirmed the Standard Model prediction of the \cenns{} cross-section.

Coherent elastic scattering of $\mathrm{\sim}$MeV reactor antineutrinos induces recoils hardly reaching the\,keV range for the lightest nuclei, requiring very low threshold detectors. 
Conventional detection techniques, only relying on the acquisition of ionization or scintillation signals, face severe challenges regarding this last point mostly because their production yield heavily quenches at low energies~\cite{lindhard1963}. 
Another key challenge is the unfavorable background conditions reactor sites usually offer. 
Those are often located at very shallow overburden ($\lesssim$ \SI{20}{\m\we}) and require sophisticated shielding strategies to achieve acceptable signal-to-background ratios. 
% Moreover, as opposed to stopped-pion sources, nuclear reactors are steady neutrino sources with rather long operating cycles and no timing information leverage other than an ON-OFF measurement for characterizing and mitigating backgrounds. 
% Despite these major experimental challenges, the detection of \cenns{} through reactor antineutrinos was recently claimed by the the CONUS+~\cite{ackermann2025} collaboration. The result was earlier claimed by the Dresden-II~\cite{colaresi2022} collaboration but has quickly been challenged by the limits set by the CONUS collaboration \cite{ackermann2024}.
% Both experiments developed very similar experimental apparatuses using kg-scale p-type point contact high purity Ge (HPGe) semiconductor detectors, deployed 10-20 m away from a commercial reactor core.
% These two experiments however showed incompatible results in the interpretation of their respective hints for a \cenns{} signal, especially regarding ionization quenching factors at sub-keV recoil energies~\cite{li2025} and reactor-correlated backgrounds~\cite{cadeddu2023}.
Moreover, as opposed to stopped-pion sources, nuclear reactors are steady neutrino sources with rather long operating cycles and no timing information leverage other than an ON--OFF measurement for characterizing and mitigating backgrounds.  
Despite these major experimental challenges, the detection of \cenns{} through reactor antineutrinos has most recently been reported by the CONUS+ collaboration~\cite{ackermann2025}. Earlier in 2022, the Dresden-II collaboration~\cite{colaresi2022} claimed the measurement but was quickly challenged by the exclusion limits obtained by CONUS~\cite{ackermann2024}.  
Both experiments developed very similar experimental apparatuses using kg-scale p-type point contact high-purity Ge (HPGe) semiconductor detectors, deployed 10--20\,m away from a commercial reactor core.
They however showed incompatible results in the interpretation of their respective hints for a \cenns{} signal, especially regarding ionization quenching factors at sub-keV recoil energies~\cite{li2025} and reactor-correlated backgrounds~\cite{cadeddu2023}.

%These two areas of tension highlight once again the experimental difficulty in achieving a reliable measurement of \cenns{} at a reactor facility.
Other reactor \cenns{} experiments using NaI scintillation detectors such as NeON~\cite{choi2023}, dual-phase time projection chambers such as RED-100~\cite{akimov2025}, Si charged coupled devices such as CONNIE~\cite{aguilar2022,aguilar2024} or HPGe semiconductor detectors such as TEXONO~\cite{kerman2025}, its successor RECODE \cite{Yang2024}, and $\nu$GeN~\cite{belov2025} are also on-going. They still miss a detection, although having acquired a significant amount of reactor-ON and reactor-OFF exposure time together with very efficient particle background rejection.
Cryogenic detection experiments, such as RICOCHET~\cite{augier2024}, MINER~\cite{agnolet2016} and \nucleus{}~\cite{angloher2019}, are also underway. 
These typically read out phonon signals induced by particle interactions in $\cal{O}$(100 g) crystal absorbers using highly sensitive thermometers and already demonstrated $\mathrm{\lesssim}$\,100\,eV energy thresholds~\cite{strauss2017,armengaud2019}.
Reaching energy thresholds below the 100\,eV range however led to the identification of another source of background, which is yet of unknown origin and commonly called the low energy excess (LEE).
The LEE is a sharp rise of events observed below a few hundreds of\,eV in many low-threshold experiments, which seems not tied to particle-induced backgrounds but rather to fundamental aspects in the design of their respective detection setups (see~e.g.~\cite{adari2022,baxter2025} for a detailed discussion and review).
The origin and the mitigation of this low-energy background component is currently the subject of many investigations and R\&D efforts, as it definitely limits sensitivity to the detection of \cenns{}~\cite{angloher2023,aguilar-arevalo2024,anthony-petersen2024,NUCLEUSlbr2025}.

The \nucleus{} experiment aims at deploying a set of ultra-low threshold \cawo{} cryogenic detectors at the Chooz nuclear power plant, which is operated by the \enquote{Électricité de France} (EdF) company. 
The experimental site dubbed the Very Near Site (VNS) is a basement room located in a tertiary building less than a 100\,m away from the two 4.25\,$\mathrm{GW_{th}}$ production units and offers a very modest overburden against secondary cosmic-ray particles~\cite{angloher2019}.
This article comprehensively reports on the characterization of the particle background environment in the VNS and on a prediction of the residual rates of particle backgrounds at sub-keV energies for the measurement of \cenns{}. 
It complements the background characterization work published by the CONUS/CONUS+~\cite{hakenmueller2019,bonet2023,sanchezgarcia2025} and RICOCHET~\cite{augier2023} \cenns{} reactor experiments, whose respective experimental sites feature more favorable overburden conditions albeit being further exposed to reactor correlated background due to their much shorter baselines.
Section~\ref{sec:nucleus_shielding} introduces the NUCLEUS experimental layout at the VNS, with a focus on the particle background shielding strategy.
Section~\ref{sec:nucleus_geant4} briefly describes the main features of the NUCLEUS simulation tool based on the \Geant{} software library, which was specifically developed for the interpretation of the different background measurements at the VNS (section~\ref{sec:vns_background}) as well as the prediction of the background rates in the cryogenic detectors (section~\ref{sec:nucleus_bck_prediction}).
Finally, section~\ref{sec:conclusion} summarizes the main results of this work and discusses future challenges for a measurement of \cenns{} at a reactor facility.

%%% ===================================================================================================================================
%%% ================================================= Section1:  NUCLEUS Shielding ====================================================
%%% ===================================================================================================================================

\section{\label{sec:nucleus_shielding}The NUCLEUS shielding system}
\begin{figure*}
	\centering
	\includegraphics[width=0.95\linewidth]{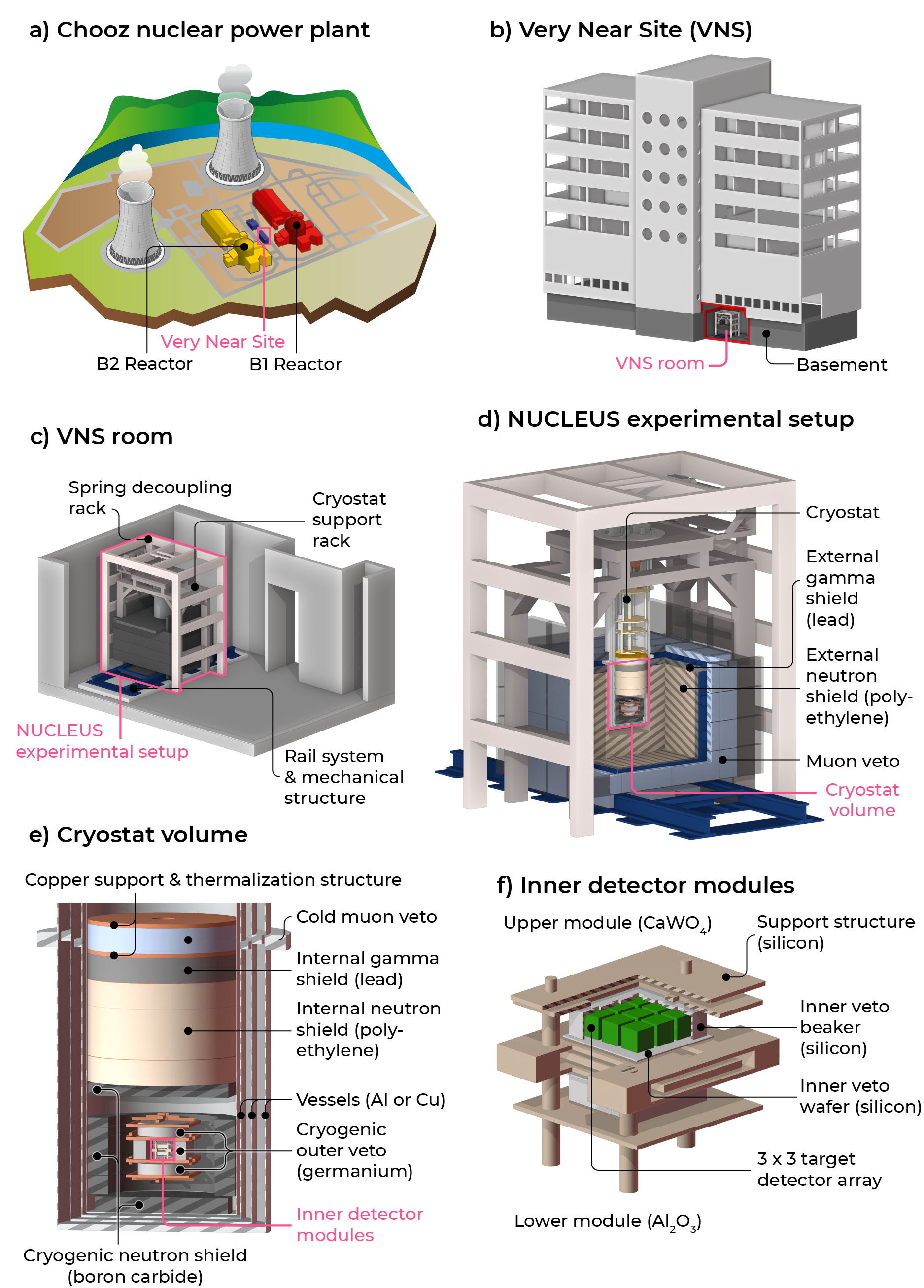}
	\caption{Simplified schematic view of NUCLEUS at the VNS, breaking down the main components of the experiment.\\
    }
	\label{fig:nucleus_layout}
\end{figure*}

Figure~\ref{fig:nucleus_layout} shows the main components of the \nucleus{} experiment at the Chooz nuclear power plant going from the inside of the setup (f) to the outside (a). 
The 10-g cryogenic detection setup (see figure~\ref{fig:nucleus_layout}f) will be operated in a dry dilution cryostat (see figure~\ref{fig:nucleus_layout}e). 
It will be exposed to a total antineutrino flux of $\mathrm{2.1\times10^{12}\,cm^{-2}\,s^{-1}}$ at the VNS location, assuming both reactors are running at nominal power.  
In the region of interest (RoI) between 10 and 100\,eV, this flux gives an average \cenns{} detection rate of about 280 and 20 \rate{} in the \cawo{} (6.8\,g) and \alo{} (4.5\,g) target detectors, respectively.
This very low rate imposes stringent requirements on the rejection of particle backgrounds, with the aim of achieving a $\cal{O}$(100) \rate{} background index in the \cenns{} RoI~\cite{angloher2019}. 
As discussed in section~\ref{sec:vns_background}, the VNS gives a very modest overburden against secondary cosmic-ray particles. 
A sophisticated shielding strategy is then required to mitigate surface muons and neutrons originating from cosmic-ray air showers, with the additional constraint of designing a small footprint setup to fit the size of the experimental room. 
%As explained later in section~\ref{subsec:neutrons_at_vns}, high energy atmospheric neutrons are the most hazardous source of backgrounds to the experiment because they are highly penetrating particles that often give the exact same experimental signature than a \cenns{} interaction.

The various NUCLEUS shielding elements are illustrated in figure~\ref{fig:nucleus_layout}d and figure~\ref{fig:nucleus_layout}e. They were designed and optimized using extensive Monte Carlo studies based on the \Geant{} simulation toolkit (see section~\ref{sec:nucleus_geant4}).
An external shielding first combines a 5-cm thick plastic scintillator-based muon veto (MV)~\cite{wagner2022} in the outermost part, together with a 5-cm thick layer of low radioactivity Pb and a 20-cm thick layer of 5\% boron-loaded high-density polyethylene (HDPE) in the innermost part.
It ensures (i) an excellent rejection of muon-induced background events and (ii) minimal production of secondary particles from muons and neutrons crossing high-Z materials while providing (iii) some attenuation to ambient gamma rays and neutrons. 
The NUCLEUS external shielding is mechanically held in place with a steel structure, which stands on a rail system, allowing it to open for easy access to the cryostat. 
It is extended within the cryostat by a so-called internal shielding (see figure~\ref{fig:nucleus_layout}e), which is housed below the still stage. 
This internal shielding follows the same layer arrangement as the external shielding, with the addition of a nearly $\mathrm{4\pi}$ 4-cm thick boron carbide (\bfourc{}) layer to further suppress neutrons reaching the target detectors. It also features a plastic scintillator-based cold muon veto, which was designed to operate at cryogenic temperatures~\cite{erhart2024}.
An essential piece of the NUCLEUS shielding system is the cryogenic outer veto (COV). 
The COV is an arrangement of two cylindrical and four rectangular 2.5~cm thick HPGe crystals mechanically held within a Cu support structure, operated at $\mathcal{O}$(mK) temperatures and read-out through the ionization channel. 
It hermetically covers the cryogenic target detectors, with the primary purpose of complementing the relatively modest attenuation of the NUCLEUS passive shielding to external gamma rays. 
It is also expected to further suppress muon-induced and neutron-induced backgrounds in the cryogenic target detectors (see section~\ref{subsec:shield_bck_reduction}). 
For the former, an energy threshold as low as possible is highly desirable. 
The concept of such an HPGe cryogenic veto system was recently validated, especially showing that a $\cal{O}$(10\,keV) threshold is within reach when operated in a dry dilution refrigerator~\cite{goupy2024}.
The last piece of the \nucleus{} shielding strategy is the target detector TES-instrumented holder, called inner veto (IV). 
The main purpose of the IV is to reject surface events and holder-related events.
It also embeds a LED system for in-situ calibration and stability monitoring of the detectors~\cite{delcastello2024}.

%%% ===================================================================================================================================
%%% ================================================ Section3:  Simulation Framework ==================================================
%%% ===================================================================================================================================

\section{Simulation framework}\label{sec:nucleus_geant4}
All simulation studies were carried out with a dedicated software using release 10.7.3 of the \Geant{} toolkit~\cite{agostinelli2003,allison2006,allison2016}.
The transport of the external background sources up to the different detection setups used throughout this work was initiated using custom-made primary event generators. These are detailed in the following corresponding sections. 
The generation of environmental gamma rays (as well as atmospheric muons and neutrons), however, shares one common point: distributing the starting positions of each primary particle along a plane tangent to a sphere (or upper-half sphere, respectively) encompassing the simulated geometry \cite{Kluck2015}.
For each Monte Carlo event, the position of the tangent plane on the sphere is randomly drawn according to the primary particle angular distribution law.
This method, later denoted as the tangent plane method, shoots primary particles perpendicularly from this plane toward the simulated geometry. 
It ensures a correct illumination of the simulated setup if the size of the tangent plane is appropriately set.
\added{For all simulations presented in this work,} particles and their interaction processes were defined using the \textit{shielding} reference physics list~\cite{allison2016}, which is particularly recommended for neutron transport. 
The electromagnetic interactions were, however, considered via the Livermore physics constructor. 
Although none of the Geant4 electromagnetic physics constructors is guaranteed to be fully reliable down to the sub-keV energy regime, the set of electromagnetic models used by the Livermore constructor is considered to be one of the most precise and realistic as it relies on an extensive library of evaluated low-energy data. 
As such, atomic relaxation processes following ionization were always switched on.
Finally, the scoring of various information of interest (energy deposition, position of interaction vertices, etc.) during particle tracking was achieved using custom-made C++ classes interfaced with the ROOT framework~\cite{brun1997} for later analysis.

As pictured in figure~\ref{fig:nucleus_layout}b to \ref{fig:nucleus_layout}f, a detailed geometry of the experimental setup and of the VNS building at Chooz was implemented for the background prediction in the NUCLEUS \cawo{} and \alo{} cryogenic detectors.
Particular attention was paid to reproducing the detector geometry with the highest possible fidelity, as the distribution of energy depositions down to the lowest energies may severely be impacted by the details of all the materials in close vicinity to the target detectors.
Additionally, specific particle production range cuts smaller than the \Geant{} default ones were applied for the target detector, inner veto, passive holder material, and cryogenic outer veto so as not to bias energy deposition in the tracking of particles in these small volumes. 
Particle production range cuts are initially expressed in units of length, and are then converted into an energy threshold beyond which \Geant{} creates and tracks new secondary particles. They are material- and particle-dependent, with an optimal value often resulting from a trade-off between computing time and the desired accuracy in the simulation of the energy deposition and stopping range of the tracked particles.
As a reasonable compromise, the target detector, inner veto, passive holder material, and cryogenic outer veto volumes were associated a particle production range cut corresponding to one-tenth of their respective smallest geometrical dimension.

Finally, time information in the particle tracking was managed using the \Geant{} stacking mechanism. 
Particle tracks occurring on timescales larger than the NUCLEUS cryogenic detector time response, i.e. $\mathrm{\sim\,100\,\mu s}$, were deferred and later processed in a next primary event.
This stacking mechanism is especially useful for correctly accounting for decaying particles and nuclei. 
A typical example is the production of unstable nuclei from muon spallation in materials near the active detectors.

%%% ===================================================================================================================================
%%% ================================================ Section4: Measurements ==================================================
%%% ===================================================================================================================================

\section{\label{sec:vns_background}Particle background environment at the VNS}
The design of the NUCLEUS shielding and the prediction of the particle backgrounds in the cryogenic detection setup required a thorough understanding of the background radiation environment at the VNS. 
This section reports on the full set of measurements performed to characterize the cosmic ray-induced muon and neutron radiations, the environmental radioactivity, airborne radon and the screening of radioactivity in the NUCLEUS detector and shielding materials. 
These measurements add up to preliminary background characterization campaigns organized at the very beginning of the NUCLEUS project and presented in~\cite{angloher2019}. 
They were used as input data to predict and normalize the different components contributing to the NUCLEUS background modeling presented in section~\ref{sec:nucleus_bck_prediction}.

%%% ================================================ Muons ==================================================
\subsection{Attenuation of cosmic ray-induced muons}\label{subsec:muons_at_vns}

The attenuation of the surface muon flux at the VNS was determined in a previous measurement campaign using a so-called \textit{cosmic wheel} apparatus, which was developed in the scope of the \enquote{Sciences à l'école} French outreach program~\cite{CosmicWheel}. 
This device consists of three parallel plastic scintillator planes read out with photomultiplier tubes (PMTs) and operated in coincidence mode to detect muons. 
The panels are mounted on a rotating frame, allowing them to point at different zenith and azimuth angles. 
Measurements of the muon count rates above ground and in the VNS at various zenith and azimuth orientations gave an omnidirectional muon attenuation factor of \SI{1.41(2)}{}, corresponding to a mean overburden of \SI{2.9(1)}{\m\we}~\cite{angloher2019}.

A \Geant{} simulation of muon transport through the VNS building (see figure~\ref{fig:nucleus_layout}b) was performed to check and corroborate these measurements. 
The starting positions of muons were distributed according to the previously described tangent plane method, using a 35-m radius sphere centered on the VNS position hence fully encompassing the concrete building together with a tangent plane of ($30\times30$)\,m$^{2}$ to fully illuminate the VNS. 
The muon initial energy and propagation direction were sampled from a modified \textit{Gaisser} parametrization adjusted to a collection of experimental data measured at various zenith angles~\cite{Tang2006}. 
The generation and tracking of ${{\sim\num{2.2e8}}}$ atmospheric muons were performed, only scoring information of primary and secondary particles reaching a 1.4-m radius spherical ghost volume, centered on the position of the NUCLEUS experimental setup in the VNS and encompassing it (see figures~\ref{fig:nucleus_layout}b) and c). 

\begin{figure*}[]
    \centering
    \includegraphics[width=0.95\textwidth]{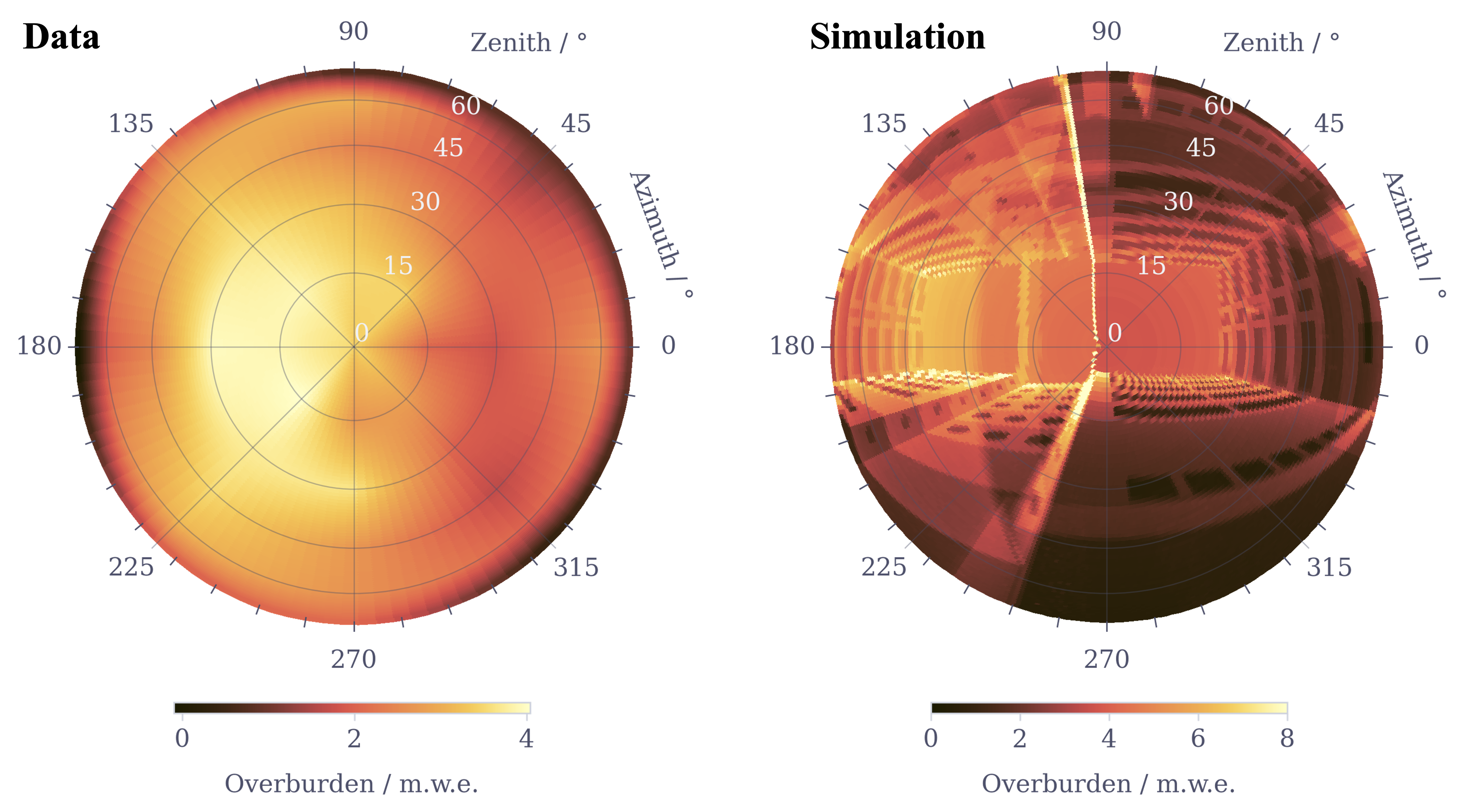}
    \caption{Projections of the VNS overburden as a function of the zenith and azimuth angles in units of meters of water equivalent (\si{\m\we}), as obtained from a measurement campaign using the \textit{cosmic wheel} detector (left) and from a \Geant{} simulation (right). See text for further details.}
    \label{fig:muons_overburden_VNS}
\end{figure*}

Figure~\ref{fig:muons_overburden_VNS} compares projections of the overburden as a function of both zenith and azimuth orientations, as obtained from the \textit{cosmic wheel} measurements and the \Geant{} simulation. 
The overburden map in the left panel of figure~\ref{fig:muons_overburden_VNS} was constructed by 2-D interpolating the set of collected measurements over a finer grid in the zenithal and azimuthal angle space.
It includes then a smearing effect resulting from the \textit{cosmic wheel}'s \SI{70}{\degree} acceptance angle to muons.
The \Geant{}-predicted overburden map is shown in the right panel of figure~\ref{fig:muons_overburden_VNS} and was built with a ($\SI{1}{\degree}\,\times\,\SI{1}{\degree}$) bin size to keep and exhibit the details of the VNS shadowing to muons when viewed in the upward direction. 
It correctly reproduces many of the features observed from the \textit{cosmic wheel} measured overburden map, such as (i) a peak attenuation toward 15-30\degree{} zenith and \SI{180}{\degree} azimuth angles, respectively and (ii) the left-right asymmetry caused by the off-centered positioning of the VNS in the underground basement of the building (see figure~\ref{fig:nucleus_layout}b). 
An omnidirectional overburden of \SI{2.92(1)}{\m\we} was computed by averaging the overburden map over all directions. 
It perfectly matches the \textit{cosmic wheel} result. 
This very good agreement validates a correct implementation of the VNS geometry and gives confidence in the description of cosmic ray-induced surface muons and their transport as currently implemented in the NUCLEUS simulation framework. 
Other features of the atmospheric muon component reaching the VNS were therefore estimated using this simulation. 
For instance, the atmospheric muon mean energy was found to be slightly shifted from 6.8\,GeV above ground to 7.6\,GeV in the VNS, further confirming the small overburden provided by the building. 
The secondary neutron and gamma total production yields from the VNS building at the position of the NUCLEUS setup were estimated via simulation to be $0.032\pm0.001$ and $1.3\pm0.1$ particle per primary muon, respectively.
When normalized to typical surface muon fluxes (see table~\ref{tab:Flux_Uncertainties}), the production of secondary neutrons is therefore expected to make a small or even negligible contribution to the measured neutron ambiance in the VNS (see section~\ref{subsec:neutrons_at_vns}). 
The same conclusion applies to muon-induced gamma rays when compared to the measured gamma-ray ambiance (see section~\ref{subsec:gamma_at_vns}), even though their production yield is a factor 40 larger than that of neutrons. 
Muon-induced gamma rays mostly come from bremsstrahlung and can reach 10-100 MeV energies, which are far larger than those of the environmental gamma-ray ambiance. 
For completeness, all muon-induced secondary particles were therefore included in the \nucleus{} background prediction (see section~\ref{sec:nucleus_bck_prediction}).

%%% ================================================ Neutrons ==================================================
\subsection{Reduction of cosmic ray-induced fast neutrons}\label{subsec:neutrons_at_vns}

Neutrons reaching the experimental setup are a particularly dangerous source of background radiation for any \cenns{} signal measurement as they are difficult to shield and able to elastically scatter off nuclei, thereby mimicking the \cenns{} experimental signature. 
In this respect, the neutron environment in the VNS deserves to be carefully and thoroughly characterized. 
A \Geant{} simulation of neutron transport through the NUCLEUS experimental setup identified fast neutrons with $\mathrm{\geq}$ 10~MeV energies in the VNS room to be the dominant source of nuclear recoils in the 10-100\,eV region of interest~\cite{GoupyPhD2024}. 
At very shallow experimental sites, such neutrons mostly originate from cosmic ray-induced, very high-energy atmospheric cascades. 
A measurement strictly focusing on characterizing the flux reduction of these very high-energy neutrons in the VNS was therefore conducted using a set of Bonner spheres.

\subsubsection{Bonner sphere measurement}\label{subsubsec:BS_measurement_at_vns}
The Bonner sphere neutron spectroscopy technique uses a set of polyethylene-based moderating spheres in combination with a thermal neutron detector to characterize a neutron ambiance over several orders of magnitude in energy~\cite{Bramblett1960}. 
The thermal neutron detector used in the present work was a 10~mm diameter and 2~mm thick $\mathrm{^{6}}$Li-enriched europium-doped lithium iodide, LiI(Eu), scintillating crystal read out by a PMT and housed in a nickel-chromium alloy case. 
This probe, designed for high flux neutron measurements, was distributed by Berthold Technologies GmBH \& Co under the part number LB6603I~\cite{Berthold}. 
The measurement of the fast neutron flux reduction at the VNS specifically used two spheres featuring very similar neutron response functions up to $\mathrm{\sim}$10~MeV: (i) a 20.32-cm diameter sphere made of high-density polyethylene (hereafter labeled BS-PE) and (ii) a 22.86-cm diameter sphere additionally including a Pb shell converter on the inside, with an outer diameter of 10.16\,cm and a thickness of 1.27~cm, to enhance the acceptance to $\mathrm{\gtrsim}$\,10\,MeV neutrons (hereafter labeled BS-w-Pb). 
Measuring the neutron capture rate difference between these two spheres combined with the LiI(Eu) probe gives access to the flux of >\,10~MeV fast neutrons.

All measurements were done by applying a 900~V bias voltage to the LiI(Eu) probe. 
The output signals were fed into an amplification chain combining a custom-made pre-amplifier and a Canberra 2012 spectroscopy amplifier. 
The amplified signals were then digitized and their pulse heights saved using an Amptek MCA 800D portable multichannel analyzer~\cite{MCA800D}. 
A calibration of the LiI(Eu) probe was attempted using $^{133}$Ba and $^{60}$Co gamma sources and a $^{252}$Cf neutron source. 
A very poor resolution was observed, probably due to a bad scintillation light collection efficiency. This, combined with the small size of the crystal relative to the mean free path of the gamma rays emitted by $^{133}$Ba and $^{60}$Co, prevented full-energy absorption peaks from being observed.
 However, the end points of the gamma spectra and the standard deviation of the $^{252}$Cf-induced neutron capture peak could be used to roughly estimate a detector energy resolution of $\mathrm{\sigma(E)~\si{[keV]}}\simeq 4\sqrt{E~\si{[keV]}}$. 
Using only the gamma sources, a $\mathrm{E=(8.5 \pm 0.5)\times i_{MCA} - (24\pm43) \,~keV_{ee}}$ linear calibration function was then estimated, where $\mathrm{i_{MCA}}$ denotes the MCA channel index, and lead to a calibration of the data in electron equivalent energy (keV$_\text{ee}$).
With such an energy calibration, the position of the \textsuperscript{6}Li neutron capture peak was reconstructed at \SI{2903(6)}{\keVee}, corresponding to a \SI{60.7(1)}{\%} quenching of the \SI{4783}{keV} total energy released by the alpha and tritium particles.
This measured quenching factor compares reasonably well with earlier measurement reported in the literature, reconstructing the neutron capture peak between 3 and 4\,MeV$_\text{ee}$~\cite{Pausch2008, Iwanowska2011, Yang2011}.

The measurement campaign at the Chooz nuclear power plant accumulated a total of 159 days of data. 
Data collection was segmented into 2-week long runs alternating each sphere to smooth temperature effects on the signal amplification chain and possible seasonal flux variations. 
Each sphere accumulated a total of 52 days of data collection at the VNS. 
In contrast, the BS-w-Pb and BS-PE spheres collected 35 and 20 days of data, respectively, when relocated to the surface.

\begin{figure*}[]
\centering
        \includegraphics[width = 1\linewidth]{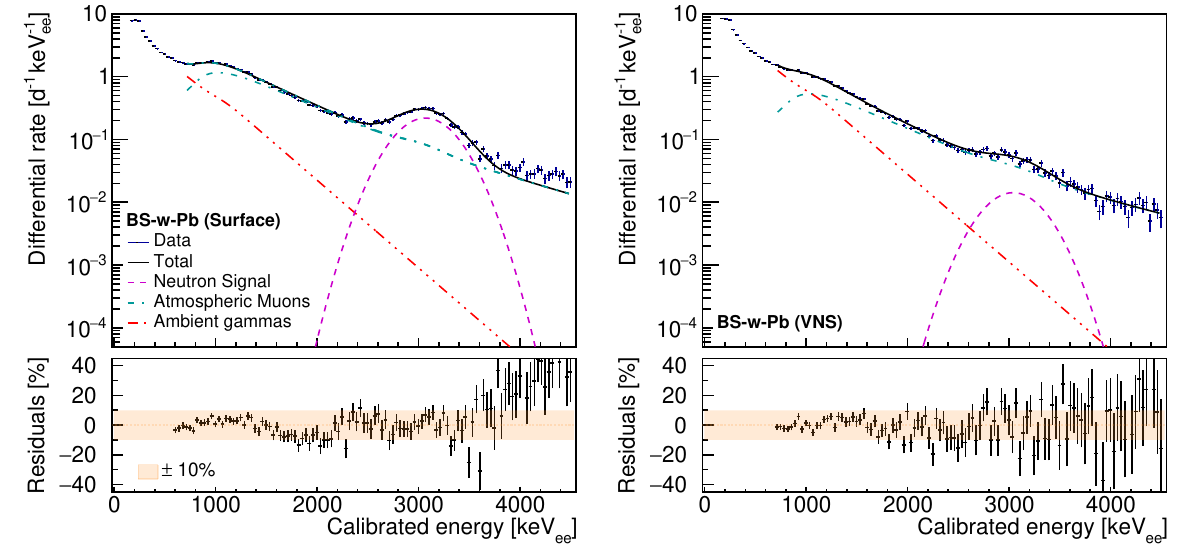}
    \caption{Best fit model of the BS-w-Pb sphere acquired data at the surface (left) and in the VNS (right) at the Chooz nuclear power plant. %With this sphere, the muon-induced background features a bump in the neutron peak region: it corresponds to the capture of muon-induced neutron through spallation in the lead layer of the sphere. 
    The bottom plots show the residuals of the fit and the $\pm$ 10\,\% band.
    }
    \label{fig:BS_fit_BS_EXT}
\end{figure*}

In the VNS, a signal-to-background ratio largely below 1 was observed, (visible it the right panel of figure~\ref{fig:BS_fit_BS_EXT}). To improve the sensitivity of the setup, both at surface and in the VNS, it was completed with a small muon veto made of a ${(\mathrm{30 \times 26 \times 3.5) \,\,cm^3}}$ plastic scintillator panel coupled to a PMT.
% Because the LiI(Eu) probe turned out to be undersized for having a measurement with a signal-to-background ratio lower than one in the VNS (see the right panel of figure~\ref{fig:BS_fit_BS_EXT}), the measurement setup was completed with a small muon veto made of a $\mathrm{30 \times 26 \times 3.5 \,\,cm^3}}$ plastic scintillator panel coupled to a PMT. 
This muon veto was placed underneath the sphere, giving a modest 30\% muon-induced background reduction in the thermal neutron capture peak region.

Because of the unfavorable signal-to-background ratio observed at the VNS, the expected backgrounds in each Bonner sphere setup had to be carefully modeled to extract the neutron capture counting rate for each measurement. 
Atmospheric muons and ambient gamma rays are the main contributors to the expected backgrounds. 
They were modeled using the NUCLEUS \Geant{} simulation framework, in the same fashion as the simulations respectively described in sections~\ref{subsec:muons_at_vns} and~\ref{subsec:gamma_at_vns}. 
These simulated background components were then linearly combined with a Gaussian model of the expected neutron capture signal to be fitted on the data~\cite{GoupyPhD2024}. 
In this work, the shape of the atmospheric muon and ambient gamma-ray contributions were assumed to be sphere-dependent but measurement location-independent (i.e. the same at the surface and in the VNS). 
The fit procedure was performed over the \SIrange{600}{4500}{\keVee} energy range, leaving the neutron Gaussian normalization, peak position and standard deviation together with the normalization of the atmospheric muon and ambient gamma-ray components as free parameters.
Figure~\ref{fig:BS_fit_BS_EXT} compares the BS-w-Pb sphere data to the best-fit model at the surface and VNS locations, showing a $\mathrm{\pm}$ 10\% agreement. A similar agreement was obtained with the BS-PE sphere.
The neutron capture rates obtained from the best fit neutron signal model of the four Bonner sphere measurements are reported in table~\ref{tab:BS_fit_result}. This table reports as well the signal-over-background ratio (S/B) computed in the neutron peak region of interest, between 2.5 and 3.9\,MeV, as well as the chi-square per degree of freedom ($\chi^2/n_{dof}$) over the whole fitting range.
% To extract the neutron capture rate, the integral of the best-fit neutron signal model was computed in the 2500-3900 $\mathrm{keV_{ee}}}$ energy range. 
% The neutron capture rates obtained from the fit between 600 and 4500~keV of the four Bonner sphere measurement runs described previously are reported in table~\ref{tab:BS_fit_result}. 
As visible from the residuals in figure~\ref{fig:BS_fit_BS_EXT}, the fit gives a worse agreement above 4\,MeV$_\text{ee}$ for surface measurements, without having a significant impact on the estimated number of counts in the neutron peaks. 

{\renewcommand{\arraystretch}{1.3}
\begin{table*}[!ht]
\caption{Neutron capture rates in the BS-w-Pb and BS-PE sphere measurement campaigns, as extracted from a fit to the data also including \Geant{}-modeled backgrounds (see text for further details). Quoted uncertainties are obtained from the fit output, only considering statistical uncertainties of the data.}
\label{tab:BS_fit_result}
\centering
\begin{tabular}{ccS[table-format=3.1(2)]S[table-format=1.2(2)]c}
\toprule
\textbf{Sphere} & \textbf{Location} & {\textbf{Neutron capture rate [d$^{-1}$]}} & \textbf{S/B} & $\boldsymbol{\chi^2/n_{dof}}$ \\
\midrule
\multirow{2}{*}{\textbf{BS-w-Pb}} & Surface & 143.4(33) & 1.34(4) & 246.9/105 \\
& VNS & 7.9(10) & 0.15(2) & 163.3/106 \\
\multirow{2}{*}{\textbf{BS-PE}} & Surface & 95.7(36) & 0.87(4) & 219.2/105 \\
& VNS & 2.1(9) & 0.04 (2) & 296.8/106 \\
\bottomrule
\end{tabular}
\end{table*}
}

The output of the fit was found to be quite sensitive to the chosen energy range. 
A systematic uncertainty was then estimated varying the lower and upper limits of the fit range. 
It was the only systematic uncertainty considered here, as other sources of systematic uncertainties were found to negligibly affect the final result. 
The difference between the extracted BS-w-Pb and BS-PE thermal neutron counting rates therefore gave a fast neutron rate above 10 MeV of ${\mathrm{47.8 \pm 5.0 \, (stat.) \pm 3.3 \, (syst.) \, d^{-1}}}$ and ${\mathrm{5.8 \pm 1.4 \, (stat.) \pm 1.3 \, (syst.) \, d^{-1}}}$ at the surface and in the VNS, respectively. 
The reduction of cosmic ray-induced fast neutrons by the VNS building was then estimated by performing a toy Monte Carlo simulation of the ratio of these two rate measurements, assuming they were normally distributed. 
The reduction factor, which was taken as the most probable value of the resulting distribution, was found to be $\mathrm{\alpha=6.8^{+3.7}_{-1.9}}$, with the quoted upper and lower values defining the 68\% highest density credible interval. 

\subsubsection{Comparison to atmospheric neutron transport simulation}
Similarly to the atmospheric muons, the measured reduction of the fast neutron flux was compared against a simulation of neutron transport through the VNS building up to the experimental area using the NUCLEUS \Geant{}-based simulation framework. 
The starting positions of neutrons were isotropically distributed using a  ${(40\,\times\,40)\,\text{m}}$ plane tangential to a 35\,m radius sphere centered on the VNS position. 
The energy of primary neutrons was sampled from a measurement of the cosmic ray-induced neutron flux and spectrum carried out on the ground~\cite{Gordon2004}, and hereafter denoted as \textit{Gordon} spectrum. 
The transport of $\mathrm{1.64\times10^{9}}$ neutrons was performed, scoring information of primary and secondary particles reaching a 1.4~m radius spherical ghost volume positioned on the NUCLEUS setup (see figures~\ref{fig:nucleus_layout}b and c). 
A total of $\sim\,3.6\times10^{6}$ neutrons was collected in the spherical ghost volume. Considering all energies and directions, this corresponds to a 1.75 reduction factor from the building. 

The left panel of figure~\ref{fig:nFluxVNSroom} compares the energy distribution of simulated neutrons reaching the NUCLEUS setup location at the VNS against the original \textit{Gordon} spectrum measured on the ground. 
They are plotted in the $\mathrm{E \times d\Phi_n/dE}$ representation, {where $\Phi_n$ is the neutron flux and ${E}$ the neutron energy, and with a log scale x-axis to better illustrate their features. 

\begin{figure*}[]
    \centering
    \includegraphics[width=\textwidth]{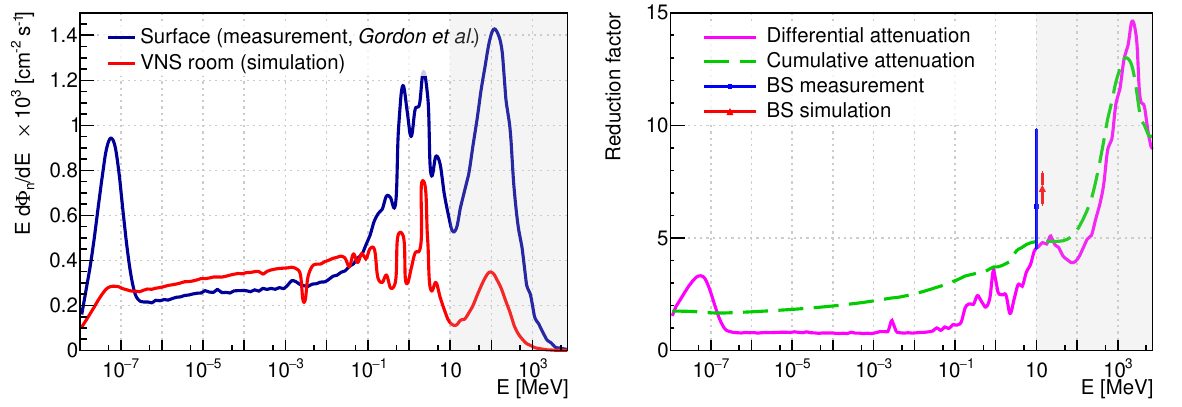}
    \caption{Reduction of the cosmic ray-induced neutron flux at the VNS. The left panel compares the \textit{Gordon} spectrum measured on ground \cite{Gordon2004} to the energy spectrum of neutrons reaching the NUCLEUS setup location as predicted by \Geant{}. The right panel displays the differential flux attenuation obtained from the ratio of these two spectra (pink solid line) and a cumulative attenuation obtained from integrating the spectra above a given energy (green dashed line). The blue square and the red triangle show the reduction factor of >10~MeV neutrons measured with the Bonner Sphere (BS) setup (see section~\ref{subsubsec:BS_measurement_at_vns}) and the >10~MeV neutron cumulative attenuation as predicted by a \Geant{} simulation of the Bonner Spheres setup, respectively. The gray areas on each panel indicate the >10~MeV energy range from which neutrons are most likely to induce a nuclear recoil in the \cenns{} RoI. See text for further details.
    }
    \label{fig:nFluxVNSroom}
\end{figure*}

The original \textit{Gordon} spectrum features two high-energy peaks typical for neutrons produced by the interaction of high-energy cosmic rays in the atmosphere. 
The highest energy peak, which is centered at $\sim$\,\SI{100}{\mega\electronvolt} and extends to GeV energies, corresponds to neutrons knocked on from atomic nuclei being collided by primary cosmic rays. 
The second peak, centered at $\sim$\,\SI{1}{\mega\electronvolt}, originates from neutron evaporation processes. 
It exhibits fine structures from nuclear resonances in the nitrogen and oxygen nuclei present in the atmosphere, but also from other nuclei in the material environment close to the neutron detection setup. 
The atmospheric neutron spectrum on the ground also features a plateau region spanning many orders of magnitude in energy between 1\,eV and 10~keV. 
This plateau corresponds to fast neutrons being moderated by the atmosphere from their production point and by the surrounding materials up to the detector. 
It finally ends with a thermal neutron peak at the lowest energies. 
The magnitude of this peak strongly depends on the surrounding materials and their composition. 
As shown on the left panel of figure~\ref{fig:nFluxVNSroom}, the VNS building is predicted to bring a sizable attenuation to the knock-on and evaporation neutrons. 
Interestingly, the magnitude of the plateau region is enhanced with respect to the surface level spectrum. 
This most likely results from the moderation of high-energy neutrons through the building materials. 
Even a dip around \SI{3}{\kilo\electronvolt} is visible and could originate from resonances in the (n,$\mathrm{\gamma}$) cross-section of $^{30}$Si, which is present in the concrete and glass compositions used to model the VNS building geometry. 
The thermal neutron peak is heavily suppressed with respect to the magnitude of the plateau region. 
The suppression of the thermal neutron peak in the simulation is likely caused by the absence of other neutron moderating materials near the ghost spherical volume. 

As shown by the right panel of figure~\ref{fig:nFluxVNSroom}, neutrons above 10\,MeV are predicted to be attenuated by a factor of approximately 5 by the VNS building.
This cumulative attenuation factor cannot, however, be directly compared to the BS setup measured attenuation because the response functions of the BS-PE and BS-w-Pb spheres slightly differ for neutron energies below 10\,MeV (see e.g.~\cite{GoupyPhD2024}). 
A \Geant{} simulation, which models the neutron response function of the two Bonner spheres to compute their expected thermal neutron capture rates, both at surface and in the VNS, predicted a total attenuation factor of ${7.2\,\pm\,0.7}$ (Monte Carlo statistical uncertainty only).
This predicted value agrees well with the previously reported BS measurement and validates the \Geant{} transport of atmospheric neutrons as implemented in the NUCLEUS simulation framework. 
The predicted differential flux attenuation (see right panel of figure~\ref{fig:nFluxVNSroom}) was therefore directly used to account for for the effect of the VNS overburden on the cosmic ray-induced neutrons in the simulation of the NUCLEUS particle backgrounds (see section~\ref{sec:nucleus_bck_prediction}).

The attenuation of cosmic ray-induced neutrons in the VNS is sensitive to the details of the implemented building geometry (e.g. material and composition). 
For the same density, changes in the composition of the concrete material used to erect the VNS building~\cite{VNS_concrete_composition} have been found to significantly influence the attenuation of low energy neutrons while having a modest impact of $\sim$\,10\% on the attenuation of neutrons above 10\,MeV.
The neutron attenuation length in concrete is $\sim$\,80\,cm above 10\,MeV and is comparable to the mean neutron path length, making a detailed implementation of the building geometry a critical aspect in the prediction of the attenuation.
For instance, a $\sim$\,10\% change in the mean neutron path length was found to change by 30\% (respectively 60\%) the total (respectively >10~MeV) attenuation to neutrons.
Since the building geometry was reproduced with a high level of fidelity with respect to the original construction drawings provided by EdF, a 30\% uncertainty on the differential attenuation to neutrons was conservatively assumed in the prediction of the \nucleus{} particle backgrounds (see table~\ref{tab:Flux_Uncertainties}). 

The energy of secondary gamma rays produced by the interaction of atmospheric neutrons in the VNS building material is predicted to range mainly between 50\,keV and 10\,MeV, with a yield of about 0.3 per primary neutron. When normalized to typical neutron fluxes as measured on the ground (see table~\ref{tab:Flux_Uncertainties}), this production yield results in a negligible (3 orders of magnitude lower) contribution with respect to the gamma-ray ambiance measured in the room (see section~\ref{subsec:gamma_at_vns}).

%%% ================================================ Gammas ==================================================
\subsection{\label{subsec:gamma_at_vns}Gamma-ray ambiance}
The gamma-ray ambiance at the VNS was characterized using a Canberra GR3018 0.68-kg n-type coaxial Reverse-Electrode Ge (REGe) detector~\cite{CanberraREGe}. 
The REGe detector is cooled down using a 7~L portable liquid nitrogen dewar, allowing a holding time of 3 to 5 days. 
The aluminum endcap is equipped with a 0.5~mm thick beryllium window at one end, which extends the detector sensitivity to low energy ($\sim$\,\SI{10}{\kilo\electronvolt}) photons. 
At the other end, a charge-sensitive preamplifier stage along with a temperature monitoring circuit are housed in an adjacent aluminum casing. 
The REGe detector was operated and data were acquired using an all-in-one DSPEC jr 2.0 module from ORTEC~\cite{DSPEC}. 
The DSPEC module amplifies, digitizes, and trapezoidally filters the analog pulses at the output of the preamplifier. 
The parameters of the trapezoidal filter, namely the rise time and the flat top width, are automatically adjusted by the MAESTRO control software~\cite{MAESTRO} for optimal energy resolution. 
This software also emulates a multichannel analyzer to measure the digital pulse amplitudes and save data in histogram form.

All the measurements reported here were acquired by applying a -5000-V bias voltage, as recommended by the REGe detector manufacturer.
The gain of the DSPEC amplifier was tuned to extend the dynamic range up to 3\,MeV, beyond the highest naturally occurring gamma-ray energy from $\mathrm{^{208}}$Tl ($\sim$\,\SI{2.6}{\mega\electronvolt}).
The detector energy response was first checked and characterized using $\mathrm{^{152}}$Eu and $\mathrm{^{60}}$Co radioactive sources, and also from a background run conducted in a surface laboratory at CEA Paris-Saclay using naturally occurring gamma-ray lines.
Excellent performances were demonstrated. A perfect linear response from $\sim$\,\SI{100}{\kilo\electronvolt} up to $\sim$\,\SI{3}{\mega\electronvolt} was observed, with e.g.~a full width at half maximum for the reconstructed 1332.5~keV $\mathrm{^{60}}$Co photo-peak of \SI{1.72}{\kilo\electronvolt}.
No stability issues were observed following multiple operations of the setup over a two-year period.
The detector energy resolution, which is of interest for the modeling of the data presented hereafter, was fitted to the following function \cite{knoll2010radiation} on the reconstruction of the most prominent naturally occurring gamma-ray lines:

\begin{ceqn}
    \setlength{\belowdisplayskip}{0pt}
    \begin{equation}
        \frac{\sigma(E)}{E} = \sqrt{\frac{A}{E^2} + \frac{B}{E} + C},
    \end{equation}
\end{ceqn}
\begin{fleqn}
    \setlength{\abovedisplayskip}{0pt}
    \begin{alignat*}{3}
        \mathrm{with\,}&\mathrm{A= (1.40 \pm 0.11)\times 10^{-1}\,keV^2,} \\
        &\mathrm{B = (2.9 \pm 0.3)\times 10^{-4}\,keV,} \\
        &\mathrm{C= (2.6 \pm 1.4)\times 10^{-9}.}
    \end{alignat*}
\end{fleqn}

Gamma-ray ambiance measurements were conducted on different sites to check the consistency of the results and the Monte Carlo interpretation of the data.
Figure~\ref{fig:Ambient_gamma_measurements} compares the ambient gamma-ray spectrum measured at the VNS together with those measured in the CEA Paris-Saclay surface laboratory and the Underground Laboratory (UGL) located at the Technical University of Munich (TU Munich).
The measured spectra all exhibit the usual naturally occurring gamma-ray lines over a continuum rising at low energies.
This continuum mostly results from the superposition of the Compton spectra associated with the many gamma-ray lines present in the surrounding environment.
The high energy portion of the spectra beyond the 2.6-MeV peak of $\mathrm{^{208}}$Tl shows the beginning of a plateau, which is mainly caused by the interactions of atmospheric muons in the detector.
A comparison of their respective levels can already give crude information about the overburden of each site.

\begin{figure}[]
    \centering
    \includegraphics[width=\linewidth]{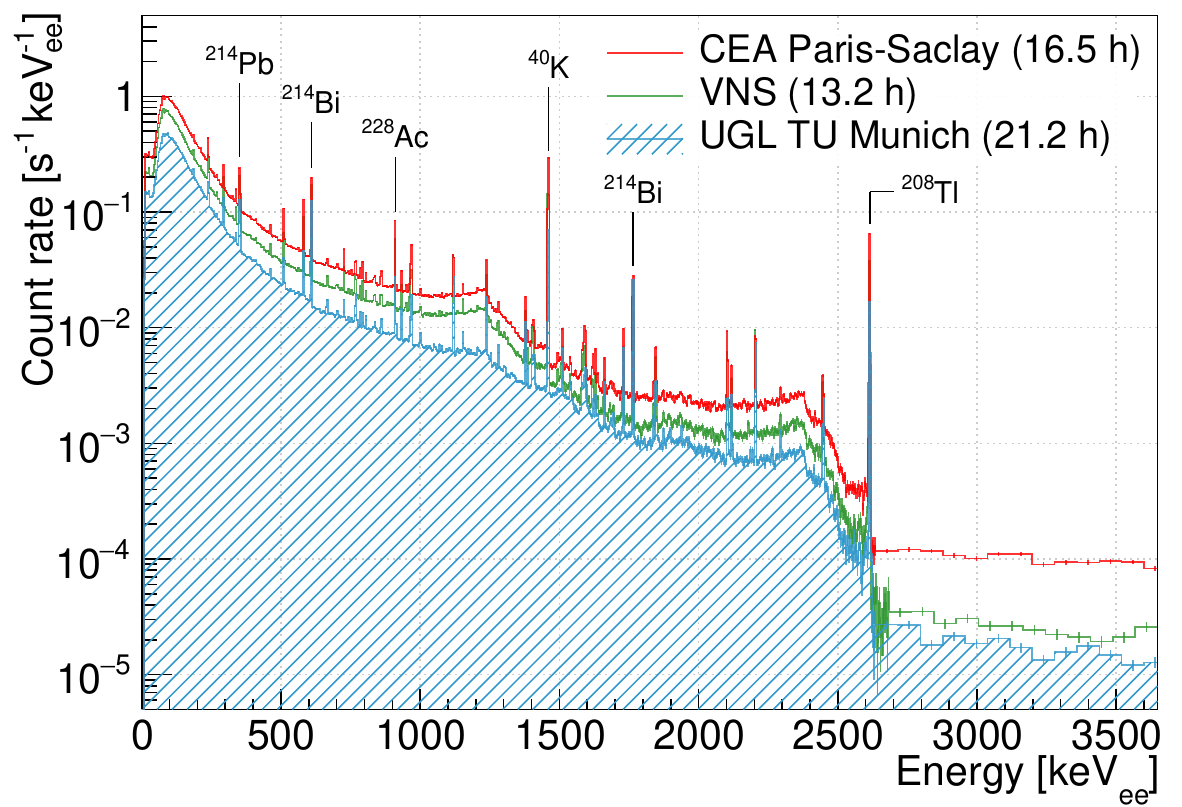}
    \caption{Gamma-ray ambiance spectra measured in the VNS (green line) and in two other \nucleus{} laboratories, the respective measuring time is stated in the legend.}
    \label{fig:Ambient_gamma_measurements}
\end{figure}

The VNS gamma-ray spectrum was extracted using a data-to-Monte Carlo comparison approach~\cite{GoupyPhD2024}.
A first step models the gamma-ray ambiance in the room by homogeneously distributing the naturally present $\mathrm{^{40}K}$, $\mathrm{^{232}Th}$ and $\mathrm{^{238}U}$ primordial radionuclides in the nearby concrete walls of the VNS building basement (see figure~\ref{fig:nucleus_layout}b), and simulating their respective decay chains assuming secular equilibrium.
The ambient gamma-ray spectrum associated with the decay of each primordial radionuclide was then constructed, saving the energy of any gamma-ray reaching a cylindrical ghost volume enclosing the REGe detector in the VNS.
In a second step, the ambient gamma-ray spectra sourcing from the decay of each primordial radionuclide decay chain were transported up to the REGe setup to predict the expected distributions of energy depositions.
A final step fitted the contributions of each of these distributions to best match the experimental data, taking into account the REGe detector energy resolution as described above.
Figure~\ref{fig:ambient_gamma_unfolding} shows the results of this procedure applied to the VNS data.
For completeness, the contributions from atmospheric muons and atmospheric neutrons were additionally simulated to achieve a better match to the data beyond the 2.6-MeV gamma-ray line of $\mathrm{^{208}Tl}$.
Their respective normalizations were fixed using the expected muon and neutron fluxes quoted in table~\ref{tab:Flux_Uncertainties}, scaling each by the expected attenuation from the VNS building.
The contribution of airborne radon measured in the room was also included (see section~\ref{subsec:radon_at_vns}). Their impact on the fit result, however, remained negligible (below 0.5\%). A good agreement at the $\pm10\%$ level is obtained between the simulation and the data over the 100-2600~$\mathrm{keV_{ee}}$ energy range. The largest deviations are observed around some gamma-ray peaks, and most probably come from small errors in the energy calibration of the data and in the energy resolution applied to the \Geant{} simulation.

The extracted contributions from $\mathrm{^{40}K}$, $\mathrm{^{232}Th}$ and $\mathrm{^{238}U}$ are reported in table~\ref{tab:primordial_contrib_to_ambient_gammas}. 
They are expressed here under units of mass activity, and are also compared to those obtained with the same method for the CEA Paris-Saclay and UGL ambient gamma-ray measurement campaigns.
They are within the range of previous natural radioactivity measurements in concrete building materials~\cite{trevisi2012natural}. 
The total ambient gamma-ray flux, which sums these three contributions, is also indicated in the last column of table~\ref{tab:primordial_contrib_to_ambient_gammas}.
%These results were cross-checked against and agreed well with another \Geant{}-based method, which subtracts the main gamma-ray line contributions from the experimental spectrum and fits the residual continuum with a combination of parametric functions.
To safely cover the remaining discrepancies between the best-fit model and the data (see bottom panel of figure~\ref{fig:ambient_gamma_unfolding}), a $\sim$\,5\% systematic uncertainty is considered additionally to the statistical uncertainties in the prediction of the NUCLEUS backgrounds sourcing from ambient gamma rays in the VNS (see section~\ref{subsubsec:bck_prediction_methods}).

{\renewcommand{\arraystretch}{1.3}
\begin{table*}[!ht]
\caption{Primordial radionuclide activities extracted from a Monte Carlo fit to the ambient gamma data measured in three different laboratory locations. The corresponding measured total gamma-ray ambiance is also indicated. Quoted uncertainties on the primordial activities are obtained from the fit output, only taking into account statistical uncertainties in the data.}
\label{tab:primordial_contrib_to_ambient_gammas}
\centering
\begin{NiceTabular}{cS[table-format=2.2(3)]S[table-format=2.2(3)]S[table-format=2.2(3)]S[table-format=2.2(3)]}
\hline
\textbf{Site} & \multicolumn{3}{c}{\textbf{Primordial activities} [Bq/kg]} & {\textbf{Total flux} [$\mathrm{cm^{-2} s^{-1}}$]} \\
& $\mathrm{^{40}K}$ & $\mathrm{^{232}Th}$ & $\mathrm{^{238}U}$ & \\
\hline
VNS   & 59.6(2) & 3.28(1) & 5.65(2) & 5.03(1)\\
UGL   & 18.33(8) & 1.91(1) & 4.89(1) & 3.12(1)\\
CEA Paris-Saclay & 82.2(4) & 5.55(1) & 7.86(1) & 7.45(1)\\
\hline
\end{NiceTabular}
\end{table*}
}
\begin{figure}[]
    \centering
    \includegraphics[width=1.\linewidth]{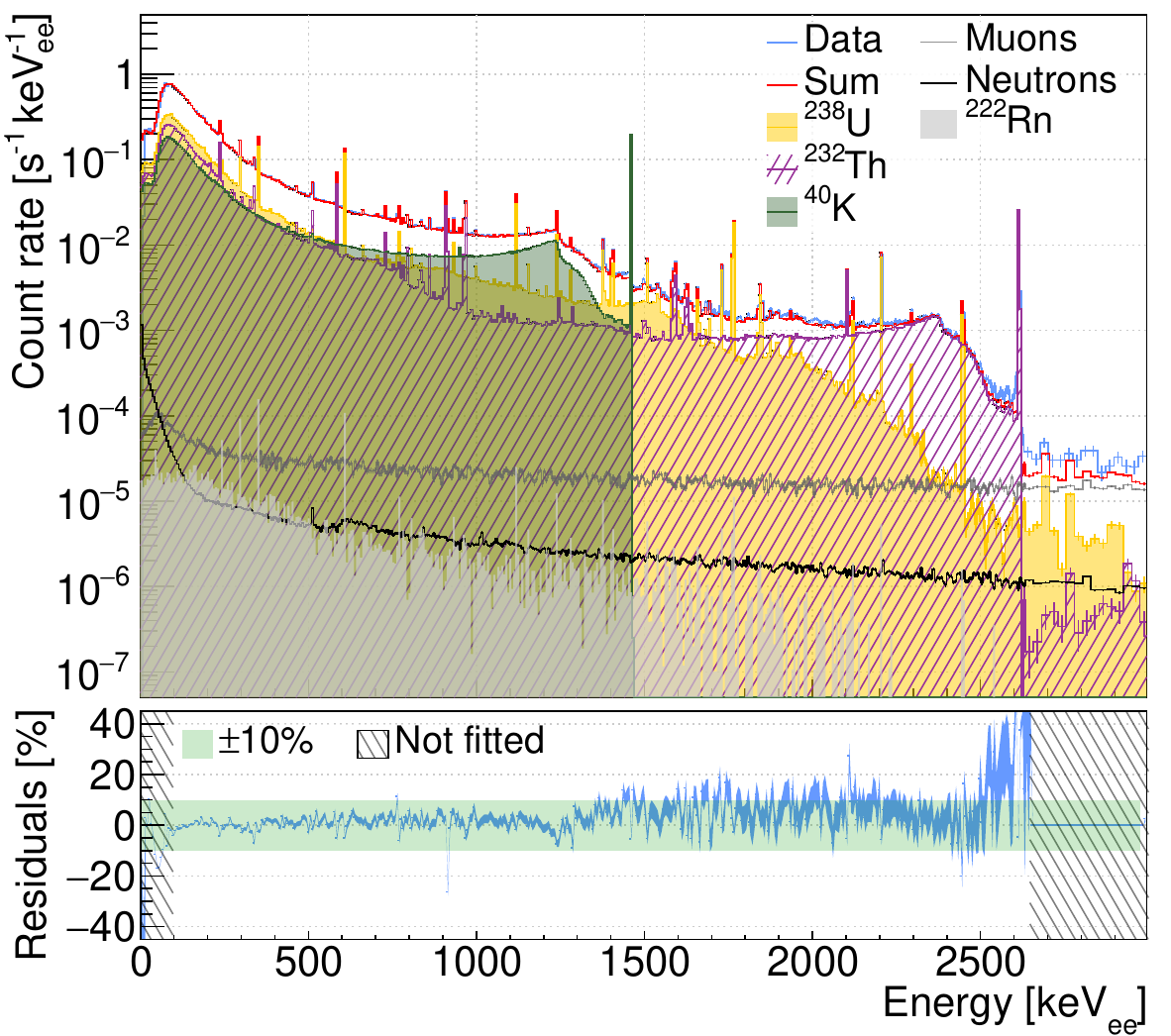}
    \caption{Extraction of the ambient gamma-ray spectrum measured at the VNS. (Top) The experimental data are matched using a \Geant{} model broken down into multiple contributions. Only the $\mathrm{^{40}K}$, $\mathrm{^{232}Th}$ and $\mathrm{^{238}U}$ primordial radionuclides contributions were fitted to the data. See text for further details. (Bottom) Residuals between the best-fit model and the data.
    }
    \label{fig:ambient_gamma_unfolding}
\end{figure}

%%% ================================================ Radon ==================================================

\subsection{Airborne radon activity}\label{subsec:radon_at_vns}
Radon is a radioactive noble gas, which is produced from the decay of primordial uranium and thorium radionuclides. 
It does not chemically bind to the materials within which its parents reside, and can therefore easily escape. 
Airborne radon can be found in high concentrations in indoor environments, especially in areas with minimal air ventilation.
The most important contribution to airborne radon comes from the $\mathrm{^{238}U}$ chain.
The fifth daughter of $\mathrm{^{238}U}$ is $\mathrm{^{226}Ra}$ and decays into $\mathrm{^{222}Rn}$. The half-life of $\mathrm{^{222}Rn}$ (3.82\,d) is long enough for much of the gas to diffuse out in the air.
Radon also comes from the $\mathrm{^{232}Th}$ and $\mathrm{^{235}U}$ decay series, which produce $\mathrm{^{220}Rn}$ (T$_{1/2}=55.6$\,s) and $\mathrm{^{219}Rn}$ (T$_{1/2}=3.96$\,s), respectively.
Because of their short half-lives, they have much greater chances to decay before becoming airborne and usually contribute much less than $\mathrm{^{222}Rn}$.
The direct progenies of radon can deposit electrostatically on any detector component surfaces in contact with air (plate-out effect), making them a significant source of background in rare event detection experiments~\cite{heusser1995}.
Particular attention must be paid to $\mathrm{^{210}Pb}$, which has a half-life of about 22 years and gives birth to the $\mathrm{^{210}Po}$ alpha emitter through two successive beta decays.
A large amount of $\mathrm{^{210}Pb}$ deposited on the detector components would then induce a background contribution affecting the whole data-taking of the \nucleus{} experiment.

Airborne radon activity in the VNS was measured %in April 2021
by using an Alpharad Plus radon monitoring device~\cite{Alpharad}.
This device uses electrostatic deposition of airborne radionuclides on a semiconductor detector operating in flow mode and was set up to collect a measurement every 23 minutes, with 3 minutes of pumping and 20 minutes of alpha particle counting.
Figure~\ref{fig:radon_at_vns} shows the original data grouped into 70-minute intervals in order to reduce statistical fluctuations. They were taken over a two-day period to assess the possible impact of human activity and air ventilation in the room. A periodic day/night variation is clearly visible.
\begin{figure}[ht!]
    \centering
    \includegraphics[width=\linewidth]{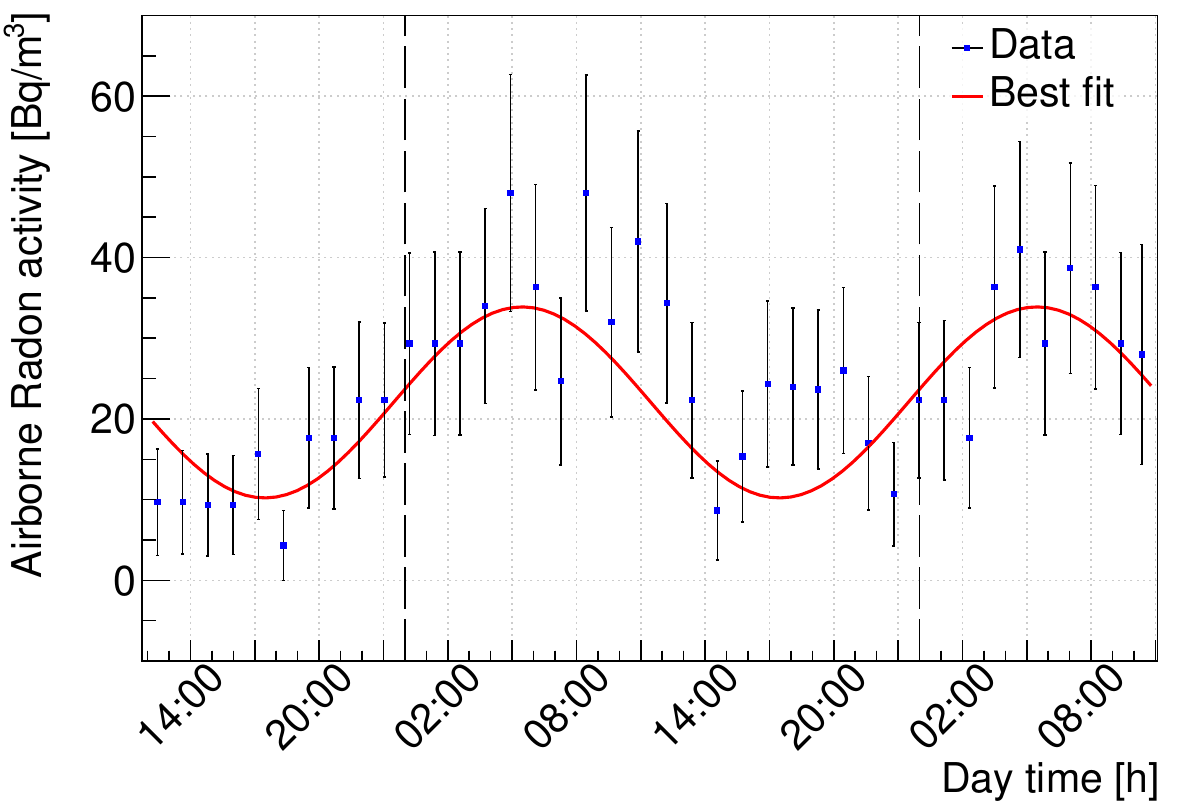}
    \caption{Measurement of airborne radioactivity in the VNS over a 2-day period. See text for further details.}
    \label{fig:radon_at_vns}
\end{figure}
Data were fitted with a sinusoidal function on top of a constant average activity. 
The period of the sinusoidal function was fixed to 1 day leaving the other parameters free. 
A constant average radon activity of \SI{22.1(1.5)}{\Bqconc} with a phase of \SI{5.5(0.7)}{h} and a sinusoidal amplitude of \SI{11.8(2.1)}{\Bqconc} were obtained, showing that radon activity levels were minimal (respectively maximal) at 6 pm (respectively 6 am) and never exceeded $\sim$\,40\,Bq\,m$^{-3}$ during the measurement. 
The origin of this modulation is yet unknown and calls for additional measurements on longer time periods.
%The VNS building being an administrative building, the daily modulation could be explained by power differences in the air conditioning system of the building between day and night. 
%
% The data points are fitted with the function: ${p0+p1 \cos \frac{2\pi}{p2} (x-p3)}$, with ${p0}$ and ${p1}$ in Bq/m$^3}$, ${p2}$ in days and ${p3}$ in hours.
% The Radon average level in the VNS room ($p0$) is $\approx}$ 20 Bq/m$^3}$ and the period ($p2$) of the cosine function is consistent with 1 day (see Figure~\ref{fig:RadonMeasurement}-\textit{left}).
% When the period ${p2}$ is kept fixed at 1 day (Figure~\ref{fig:RadonMeasurement}-\textit{right}), a phase ${p3=(5.47\pm0.67)}$ hours is obtained.
% In conclusion, a day/night effect seems to be present with a maximum Radon concentration around 6\,a.m. and a minimum around 6\,p.m.
% even at its maximum level, the Radon concentration in the VNS room does not exceed $\approx}$ 40 Bq/m$^3$.
%

Thanks to the presence of the air ventilation system, airborne radon activity in the VNS was found to be well below typical activities expected in a basement room (200-600~Bq~m$^{-3}$).
As shown in figure~\ref{fig:ambient_gamma_unfolding}, it is expected to contribute less than 0.01\% to the total gamma-ray ambiance. 
It can be then neglected as an external source of background, even in the inside of the \nucleus{} passive shield where the environmental gamma rays are attenuated by more than a factor 50 (see section~\ref{subsubsec:gammas_mitigation}).
Finally, the possible plate-out effect on the setup components, especially those located inside the cryostat in close vicinity to the cryogenic detectors, were neglected in the present work because (i) the material screening measurements presented in table~\ref{tab:Screening_result} include this contribution and (ii) most of a cumulated surface contamination can be mitigated by a simple cleaning~\cite{bruenner2021}.

%was estimated to contribute less than a milli count per day (mcpd) in the \nucleus{} array of \cawo{} detectors between 10 and 100\,eV. 
%This contribution is much smaller than the other background components (see table~\ref{tab:bckg_budget}), making airborne radon also negligible in this respect.

%%% ================================================ Screening ==================================================
\subsection{Material radiopurity}\label{subsec:material_radiopurity}
Radioactivity intrinsically present in the materials composing the experimental setup can be a key consideration in rare event search experiments. 
This radioactivity can either be of primordial ($^{232}$Th, $^{238}$U, $^{235}$U, $^{40}$K and their progenies), cosmogenic (i.e. cosmic ray activation of materials) or anthropogenic (e.g. $^{60}$Co and $^{137}$Cs) origin.
The decay of radioactive isotopes surrounding the detection volumes can produce background counts in the \cenns{} RoI. 
As such, most materials ranging from the outermost up to the innermost parts of the \nucleus{} setup were screened using non-destructive gamma spectrometry measurements at the \textsc{Stella} (SubTErranean Low Level Assay) facility of the Laboratorio Nazionale del Gran Sasso (LNGS). The \textsc{Stella} facility comprises 15 HPGe detectors sensitive to radioactivity levels as low as a few \si{\micro\becquerel\per\kilo\gram} in the screening of material samples~\cite{laubenstein2017}.

{\renewcommand{\arraystretch}{1.3}
\begin{sidewaystable}[]
    \caption{Material screening measurements of the main \nucleus{} setup components. Reported activities are expressed in \si{\milli\becquerel\per\kilo\gram}. Different progenies originating from the decay chain of the $^{238}$U and $^{232}$Th primordial radionuclides were measured to check their respective secular equilibrium conditions. Uncertainties are given at the 68\% confidence level. Upper limits are given at the 90\% confidence level.}
    \label{tab:Screening_result}
    \centering
    \footnotesize
    \begin{tabular}{llllllllll}
    \toprule
    &&& \multicolumn{2}{c}{\textbf{$^{232}$Th series}} & & &\multicolumn{3}{c}{\textbf{$^{238}$U series}}   \\ \cmidrule{4-5} \cmidrule{8-10}
    \textbf{Component} & \textbf{Mass} & \textbf{Material} & \textbf{$^{228}$Ra} & \textbf{$^{228}$Th} & \textbf{$^{40}$K} & \textbf{$^{235}$U} & \textbf{$^{238}$U} & \textbf{$^{226}$Ra} & \textbf{$^{210}$Pb} \\ \midrule
    Mechanical Structure & 1428.5\,kg & Steel & <\,16 & 6\,$\pm$\,3 & <\,98 & <\,3.4 & <\,170 & 12\,$\pm$\,3 \\
    External gamma shield & 2417\,kg & Low radioactivity lead & <\,16 & <\,13 & <\,95 & <\,32 & <\,270 & <\,2.8 & (89\,$\pm$\,7)\,$\times10^3$\\
    External neutron shield & 450.3\,kg & 5\,\%-Borated polyethylene & 1.0\,$\pm$\,0.5 & 1.5\,$\pm$\,0.3 & 13\,$\pm$\,4 & <\,0.73 & <\,21 & 2.7\,$\pm$\,0.4\\
    Vacuum vessel & 12\,kg & Al6061  & 40\,$\pm$\,14 & 370\,$\pm$\,30 & <\,160 & 130\,$\pm$\,40 & 3500\,$\pm$\,800 & <\,8.3 \\
    50K vessel  & 3.35\,kg & \multirow{2}{*}{Al1100} & \multirow{2}{*}{120\,$\pm$\,60} & \multirow{2}{*}{320\,$\pm$\,50} & \multirow{2}{*}{<\,520} & \multirow{2}{*}{500\,$\pm$\,100} & \multirow{2}{*}{(8\,$\pm$\,2)\,$\times10^3$} & \multirow{2}{*}{<\,42}\\
    4K vessel & 4.51\,kg \\
    Still vessel & 2.11\,kg &  Gold-plated copper & <\,14 & <\,14 & <\,130 & <\,6.2 & <\,76 & <\,14 \\
    Internal gamma shield  & 39.3\,kg & Low radioactivity lead & <\,15 & <\,14 & <\,71 & <\,29 & <\,310 & <\,10.6 & (40\,$\pm$\,4)\,$\times10^3$\\
    Internal neutron shield & 14\,kg & 5\,\%-Borated polyethylene & 1.0\,$\pm$\,0.5 & 1.5\,$\pm$\,0.3 & 13\,$\pm$\,4 & <\,0.73 & <\,21 & 2.7\,$\pm$\,0.4\\
    Cryogenic neutron shield & 20.90\,kg & Boron carbide (B$_4$C) & 240\,$\pm$\,30 & 460\,$\pm$\,40 & <\,360 & 30\,$\pm$\,10 & 800\,$\pm$\,300 & 260\,$\pm$\,20 \\
    Detector housing & 1.51\,kg & Copper & <\,3.3 & <\,4.1 & <\,23 & <\,3.8 & <\,92 & <\,3.5 & \\
    Detector clamps & 0.15\,g & Bronze & <\,60 & <\,26 & <\,510 & <\,15 & <\,750 & <\,40 & (4.2\,$\pm$\,0.9)\,$\times10^3$\\
    Black paint & $\sim$\,20\,g & Acrylic & 6240\,$\pm$\,380 & 5920\,$\pm$\,370 & 4900 \,$\pm$\,600 & 440\,$\pm$\,90 & 7000\,$\pm$\,2000 & 5340\,$\pm$\,220 &  \\
     Target detectors & 6.82\,g & Calcium tungstate (CaWO$_4$) & 0.034\,$\pm$\,0.003  & 0.024\,$\pm$\,0.007  & <\,0.01  & 0.3\,$\pm$\,0.01  &  3.1\,$\pm$\,0.02 & 0.174\,$\pm$\,0.006 & 0.11\,$\pm$\,0.006  \\
%    Detectors PCBs ($\times10^3$) & 8.6\,$\pm$\,0table.5 & 9.0\,$\pm$\,0.5 & 5.9\,$\pm$\,0.6 & 0.35\,$\pm$\,0.06 & 7\,$\pm$\,1 & 5.5\,$\pm$\,0.2\\
    \bottomrule
    \end{tabular}
\end{sidewaystable}
}

Table~\ref{tab:Screening_result} summarizes the results of the screening measurement campaign. 
The neutron cryogenic (\bfourc{}), gamma (Pb) and neutron (5\%-borated HDPE) shields as well as the cryostat vessels show the largest activities among the screened \nucleus{} setup components. 
In particular, the \bfourc{} cryogenic shielding closely surrounds the cryogenic detectors and is therefore expected to make a sizable contribution to the background. 
Table~\ref{tab:Screening_result} also reports the screening result of a black paint, which currently coats the inner surface of the radiation shields surrounding the COV (see figure~\ref{fig:nucleus_layout}e). 
This black paint mitigates infrared reflections, which can otherwise lead to a fast degradation of the COV performances~\cite{GoupyPhD2024}. 
Although this acrylic paint shows the highest levels of radioactivity, it is expected to have little impact on the background budget (see section~\ref{subsubsec:material_radioactovity_mitigation}) as it is used in very small amounts. However, other infrared radiation-proof solutions are currently being investigated to avoid the use of such paint. 

Some components used in the \nucleus{} setup still remain to be screened to draw a comprehensive background budget from material contamination. 
Among them, the intrinsic radioactivity of the target detectors and of the COV HPGe crystals needs to be assessed. 
Silicon materials and connectors used in the mounting of the cryogenic detection setup (see figure~\ref{fig:nucleus_layout}f), such as e.g. the IV, silicon support structure, cabling, and LED fibers remain to be characterized. 
For the background estimate presented in~\ref{subsubsec:material_radioactovity_mitigation}, only the target detectors were considered among the above mentioned non-screened material components, with radioactivity levels estimated from~\cite{CaWO4contlevels}.

%%% ===================================================================================================================================
%%% ================================================ Section5: Background prediction ==================================================
%%% ===================================================================================================================================

\section{Prediction of particle backgrounds in the \nucleus{} detectors}\label{sec:nucleus_bck_prediction}

%%% ================================================ Method ==================================================
\subsection{Assumptions and methods}\label{subsubsec:bck_prediction_methods}
%This section deals with predicting the residual background rates and spectra expected in the \nucleus{} cryogenic target detectors at sub-keV energies. 
The prediction of the residual background rates and spectra in the \nucleus{} cryogenic target detectors uses the characterization of the VNS particle background environment reported in section~\ref{sec:vns_background} as an input to the \Geant{}-based simulation framework described in section~\ref{sec:nucleus_geant4}.
Four background components are simulated: (i) atmospheric muons and (ii) atmospheric neutrons (i.e.~cosmic ray-induced background radiations), (iii) environmental gamma rays and (iv) intrinsic radioactivity coming from the \nucleus{} setup materials (see also table~\ref{tab:Flux_Uncertainties}).
Among the characterized background sources, only airborne radon was disregarded (see related discussion in section~\ref{subsec:radon_at_vns}).

In the following work, the simulation-generated datasets were analyzed by applying combinations of various selection criteria in order to identify and reject background-like events. 
First, an event is selected if at least one of the target detectors shows a hit, where a detector hit is defined as an energy deposition greater than a given threshold.
In a second step, the impact of the \nucleus{} veto detectors can be studied by further requiring no hit in one or a combination of them.
The energy thresholds defining a MV hit, a COV hit, an IV hit and a cryogenic target detector hit were set to 5 MeV, 1\,keV, 30\,eV and 10\,eV, respectively. 
Finally, the identification of a \cenns{}-like event must meet the combination of all possible anti-coincidence selection criteria, i.e. having (i) no hits in any of the veto detectors and (ii) a hit in one and only one of the target detectors. 
In the following work, the time and energy response of all the active detectors were disregarded.
Although the \nucleus{} simulation analysis framework allows to predict the fraction of events surviving a set of selection criteria, assumptions about the normalization and associated uncertainty for each of the considered background components are necessary to estimate absolute rates and spectra. 
These are listed in table~\ref{tab:Flux_Uncertainties} and discussed in the following. Since they are normalization uncertainties they are neglected for the relative shielding attenuation studies presented in section~\ref{subsec:shield_bck_reduction} and they are only propagated to the residual background rate in the VNS presented in section~\ref{sec:vns_background}.
\begin{table*}[ht!]
    \caption{Flux and uncertainty assumptions used in the normalization of the different background components. The quoted uncertainties only apply to the computation of the expected rates and spectra of background events at the VNS. See text for further details.}
    \label{tab:Flux_Uncertainties}
    \renewcommand{\arraystretch}{1.3}
    \centering
    \begin{NiceTabular}{lcc}
        \toprule
        \textbf{Background component} & \textbf{Flux [\si{\per\square\centi\meter\per\second}]} & \textbf{Uncertainty (VNS) [\%]} \\ \midrule
        Atm. muons & 1.90 $\times\,10^{-2}$ (surface) & 25 \\
        Atm. neutrons & 1.34 $\times\,10^{-2}$ (surface) & 30 \\
        Env. gamma rays & 5.03 (VNS) & 20 \\
        Material radioactivity & see table~\ref{tab:Screening_result} & 30 \\
        \bottomrule
    \end{NiceTabular}
\end{table*}

The normalization of the cosmic ray-related neutron and muon components are based on literature values, as an absolute flux measurement at the VNS is at this stage missing for both. 
The flux value of the muon component comes from integrating the modified \textit{Gaisser} parametrization, with an uncertainty mostly driven by the experimental measurements it is adjusted on~\cite{Tang2006}, but also including unaccounted-for effects such as the impact of atmospheric pressure changes and solar modulation of the surface muon flux.
The neutron component is normalized using the \textit{Gordon} flux measurement \cite{Gordon2004}. 
%This measurement was found to be consistent at the 10-20\% level with other flux measurements reported in the literature \cite{CONUS+background, Pioch2011, Goldhagen2003}, especially in the >10 MeV range. 
The quoted uncertainty includes uncertainties from (i) spread in similar measurements published in the literature (see e.g.~\cite{sanchezgarcia2025,Pioch2011,Goldhagen2003}) and (ii) the predicted attenuation from the VNS building (see related discussion in section~\ref{subsec:neutrons_at_vns}). 
The ambient gamma-ray component is normalized using the measurement described in section~\ref{subsec:gamma_at_vns}. 
Although a systematic uncertainty of 5\% was assessed, an overall 20\% uncertainty was assumed to safely cover any changes in the gamma-ray ambiance following the introduction of auxiliary equipments and materials other than the screened \nucleus{} components as listed in table~\ref{tab:Screening_result}. 
The last background component is internal to the \nucleus{} setup and adds up the primordial radionuclide activity of each of the screened materials (see section~\ref{subsubsec:material_radioactovity_mitigation}). The quoted uncertainty applies to the sum of all contributions, and comes from a crude but conservative estimate, which combines (i) the statistical uncertainty of the least radio-pure material screening measurement (i.e. the cryogenic neutron shield $\mathrm{B_4C}$), (ii) MC statistical uncertainties and (iii) the difference between modeling the material radioactivity component either setting reported upper limits in table~\ref{tab:Screening_result} to zero or using them as central values.

The present modeling is not intended to provide a fine estimate of the expected background data, but rather to detail the \nucleus{}  particle background mitigation strategy and to give a range of background levels that can reasonably be expected at the Chooz nuclear power plant.
Most importantly, it lays the foundations of a future model, which will be validated and fine tuned using the \nucleus{} commissioning data~\cite{NUCLEUSlbr2025} and future data taken at the VNS,~such that a potential \cenns{} signal can be confidently and robustly extracted.
The following two sections respectively discuss the performances of the \nucleus{} shielding system for the mitigation of particle backgrounds and the particle background budget expected at the \nucleus{} experimental site.

%%% ================================================ NUCLEUS shielding mitigation ==================================================
\subsection{Background reduction from the NUCLEUS shielding system}\label{subsec:shield_bck_reduction}
\subsubsection{Mitigation of cosmic ray-induced radiations}\label{subsubsec:CR_mitigation}
%
%%%%%%%%%%%%%%%%%%%%%
%%% ATM. NEUTRONS %%%
%%%%%%%%%%%%%%%%%%%%%
%
As already mentioned in section~\ref{sec:nucleus_shielding}, the mitigation of cosmic ray-induced backgrounds is a paramount requirement in \nucleus{} because (i) the VNS is a very shallow site and (ii) cosmic-ray radiations unavoidably lead to the production of secondary particles, especially neutrons, when interacting in the materials surrounding the cryogenic target detectors. 
The shielding performances to mitigate surface muons and neutrons were evaluated using a so-called open-air geometry, which ignores the VNS building. 
The starting position and energy of the primary particles were drawn using the atmospheric muon and neutron custom event generators already described in sections~\ref{subsec:muons_at_vns} and~\ref{subsec:neutrons_at_vns}, respectively. 
The size of the tangent plane for illuminating the open-air geometry was smaller than the one used for the VNS building attenuation studies. It was optimized to $120\times120$\,cm$^{2}$ (respectively $280\times280$\,cm$^{2}$) for muons (respectively neutrons). 
For the case of atmospheric neutrons, a crude but conservative approximation scaled down all deposited energies in the COV and the MV volumes by a factor 5 and 2 respectively, to take into account their quenching to neutron-induced nuclear recoils \cite{Bonhomme2022, Laplace2020, Awe2021}. For muon-induced background, since in the COV, most of the events seen in coincidence with the cryodetectors come from secondary neutrons, a factor 5 down scaling of the energies deposited in the COV is cautiously applied as well.
% The analysis of the simulation data applied various combinations of selection criteria to investigate the impact of the different \nucleus{} veto systems. 
% The computation of the expected rates of surviving events used the flux values quoted in table~\ref{tab:Flux_Uncertainties}.

The top panels of figure~\ref{fig:CR-induced_shielding_mitigation} display the rate of energy deposition between 0 and 1\,keV in the \cawo{} array of target detectors
induced by the atmospheric neutron component.
The top left panel shows how it is reduced by the successive addition of the different \nucleus{} passive shielding layers. 
In a configuration where no shielding is present around the detection setup, the predicted ${10^{4}-10^{5}}$\,\rate{} event rate in the RoI is far greater than the expected \cenns{} signal and confirms the need for adding neutron shields to moderate and attenuate this background component.

\begin{figure*}[ht!]
    \centering
    \includegraphics[width=1\textwidth]{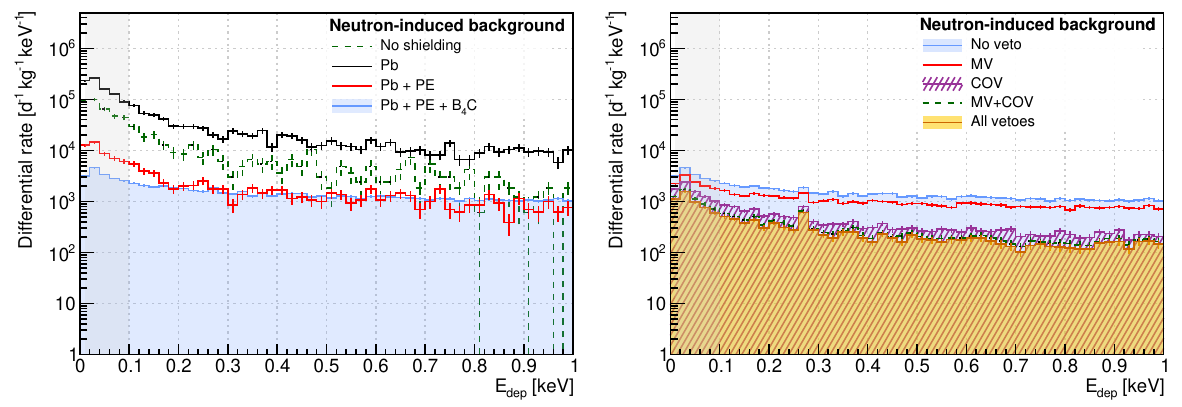}
    \includegraphics[width=1\textwidth]{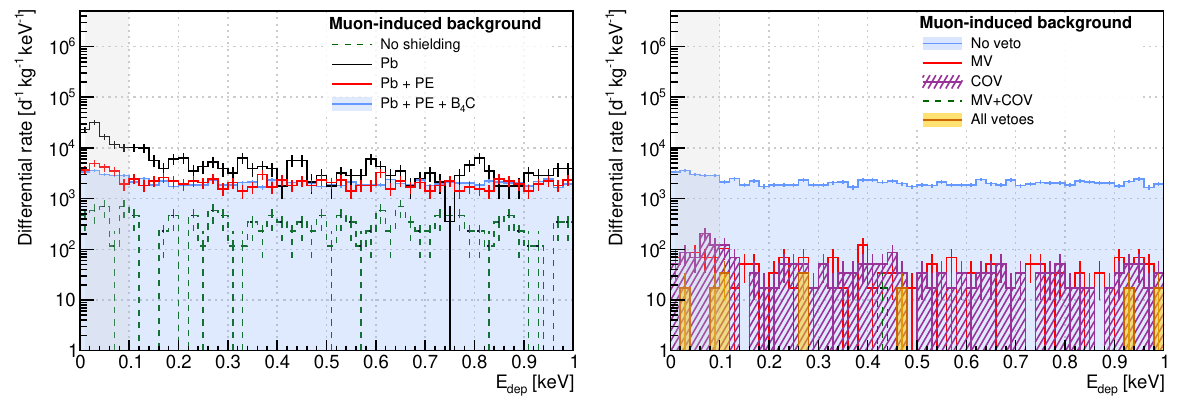}
    \caption{Reduction of the cosmic ray-induced backgrounds by the \nucleus{} shielding, at the surface. The top (respectively bottom) panels deal with the atmospheric neutron (respectively atmospheric muon) component. Each histogram shows the rate of events with deposited energy between 0 and 1\,keV in the \cawo{} array of target detectors. Uncertainties are statistical only. The left panels show the impact of sequentially adding passive shielding layers. The right panels show how using the different veto detectors complements the passive shields. The ``all vetoes'' selection criteria apply all possible anti-coincidence criteria for the rejection of background events. The \cenns{} RoI is indicated by the light gray filled area.}
    \label{fig:CR-induced_shielding_mitigation}
\end{figure*}

Because of spallation and ${(n, xn)}$ reactions of >10 MeV neutrons in Pb, the external and internal (i.e. inside the cryostat, see figure~\ref{fig:nucleus_layout}e) gamma shields are detrimental to the suppression of neutron-induced backgrounds, strengthening the need for adding neutron shields.
The addition of a first 20-cm thick borated HDPE layer reduces by more than one order of magnitude the residual event rate in the \cenns{} RoI. 
A second 4-cm thick $\mathrm{B_4C}$ layer, which is located within the cryostat in the direct vicinity of the cryogenic detection setup (see figure~\ref{fig:nucleus_layout}e), captures the low-energy ($\lesssim$\,10\,keV) component of the surviving neutron flux. 
It turns out to be a very effective complement to the external neutron shield, further suppressing the event rates in the \cawo{} detectors by a factor $\sim$\,5. 
The emission of a 478\,keV gamma-ray line following neutron capture on $\mathrm{^{10}B}$ is harmless to the cryogenic target detectors, as it is efficiently shielded and vetoed by the 2.5\,cm thick COV HPGe crystals.
The top right panel of figure~\ref{fig:CR-induced_shielding_mitigation} shows the impact of the different veto detectors in addition to the attenuation of the passive shield layers. 
While the impact of the MV is marginal, the COV brings a sizable additional reduction of the neutron-induced backgrounds of a factor 5. 
The COV rejection performances particularly depends on its ability to detect low energy nuclear recoils induced by 10-100\,keV neutrons reaching the target detectors, as those are mostly driving the rate of residual background events in the \cenns{} RoI~\cite{GoupyPhD2024}. 
For instance, raising the COV threshold from 1 $\mathrm{keV_{ee}}$ to 10 $\mathrm{keV_{ee}}$ degrades the neutron-induced background rejection by approximately 20\%. 
Finally, the benefit of applying all vetoes, i.e. (i) using the IV and (ii) requesting only one cryogenic detector hit in addition to the MV and COV anti-coincidence selection criteria, is very marginal. 
This result makes sense as the IV and the gram-scale cryogenic target detectors are small compared to the mean free path of\,keV to MeV neutrons either in Si, \cawo{} or \alo{} materials.

%
%
%%%%%%%%%%%%%%%%%%%%%
%%% ATM. MUONS %%%%%%
%%%%%%%%%%%%%%%%%%%%%
%

In a similar way, the bottom panels of figure~\ref{fig:CR-induced_shielding_mitigation} comprehensively details the \nucleus{} shielding mitigation of surface muon-induced backgrounds. 
The bottom left panel first confirms the importance of keeping to a minimum the amount of high-Z materials for limiting the production of muon-induced secondary particles \cite{Kluck2015}.
In this regard, a thickness of 5\,cm of Pb was chosen for both the external and internal gamma shields such that sufficient suppression of external gamma rays can still be achieved when it is combined with the COV (see section~\ref{subsubsec:gammas_mitigation}).
The bottom left panel of figure~\ref{fig:CR-induced_shielding_mitigation} also shows that the addition of neutron shields within the Pb layer reduces the background rate in the \cawo{} array of target detectors mostly at very low deposited energies. 
This can be interpreted as the fact that $\sim$\,MeV neutrons produced by muon interactions in Pb are moderated to 1-100\,keV energies and partly attenuated by the HDPE and \bfourc{} layers. When reaching the target detectors, they then elastically scatter off tungsten nuclei, which predominantly recoil with energies in the $\mathcal{O}$(100\,eV) regime. 
The interaction of muons in Pb produces gamma rays in greater quantities than neutrons, mostly through bremsstrahlung processes~\cite{groom2001}. 
This fact both explains why (i) as previously shown on the bottom left panel of figure~\ref{fig:CR-induced_shielding_mitigation}, the addition of neutron shield layers makes little impact to the background rates and spectra and (ii) the COV is very efficient in reducing the muon-induced backgrounds, as shown on the right panel of figure~\ref{fig:CR-induced_shielding_mitigation}.
Unsurprisingly, the MV was also found to be very efficient. When combined with the COV, they are predicted to reject more than 99.8\% of the muon-induced backgrounds in the \cenns{} RoI, making them a negligible contributor to the total background budget (see section~\ref{subsec:bck_budget}).
%
%
% \begin{itemize} 
%     \item Properties of the primary muons that produce background in ROI
%     \item Effect of the OV and MV energy thresholds
%     \item Expected counting rates and spectra in the target detectors (for different applied cuts)
% \end{itemize}
%
%
%%%%%%%%%%%%%%%%%%%%%
%%% ENV. GAMMAS %%%%%
%%%%%%%%%%%%%%%%%%%%%
%
\subsubsection{Mitigation of environmental gamma rays \label{subsubsec:gammas_mitigation}}

% \begin{itemize}
%     \item Energy of the ambient gammas that can produce background in ROI
%     \item Effect of the OV energy threshold
%     \item Expected counting rates and spectra in the target detectors (for different applied cuts)
% \end{itemize}
The mitigation of environmental gamma rays by the \nucleus{} shielding was studied using the best fit modeling of the gamma-ray ambiance measured in the VNS room (see section~\ref{subsec:gamma_at_vns}).
Ambient gamma rays were transported up to the cryogenic detectors, with starting positions isotropically distributed using a ${120\times120}$\,cm$^{2}$ plane tangential to a 3-m radius sphere centered on the open-air geometry.
% Similarly to the cosmic ray-induced backgrounds, the analysis of the simulation data applied various combinations of selection
% criteria to investigate the impact of the different NUCLEUS veto systems. 
In the following results, the computation of the expected rates of surviving events used the absolute ambient gamma-ray flux measurement described in section~\ref{subsec:gamma_at_vns} and quoted in table~\ref{tab:Flux_Uncertainties}.

Figure~\ref{fig:Gamma-induced_shielding_mitigation} displays energy deposition spectra obtained between 0 and 1\,keV in the \cawo{} array of target detectors. 
The left panel illustrates how the \nucleus{} experimental setup passive materials contribute to the suppression of the ambient gamma-ray background component, especially focusing on the Pb external gamma shield. 
In the absence of any shielding, the measured gamma ray ambiance at the VNS results in a flat distribution of events with a rate of $\sim10^{4}$\,\rate{}. 

\begin{figure*}[ht!]
    \centering
    \includegraphics[width=1.\textwidth]{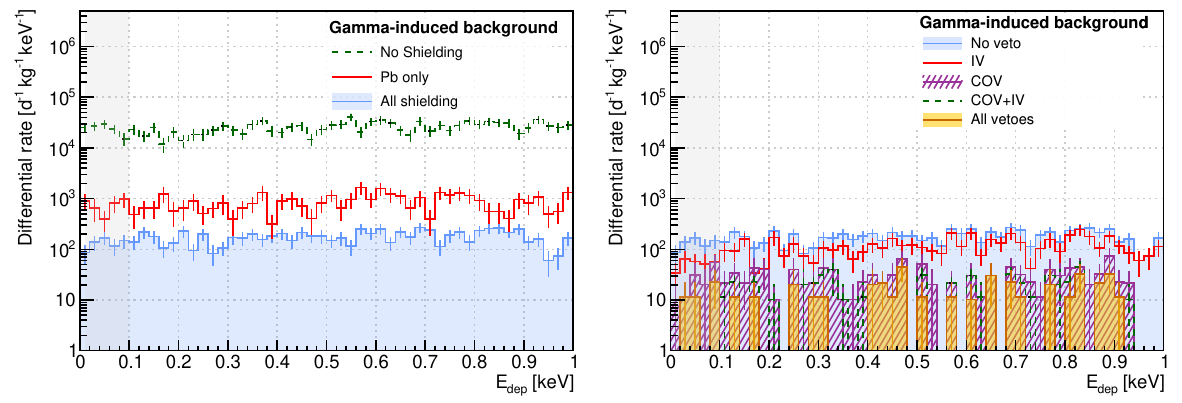}
    \caption{Reduction of the ambient gamma ray-induced backgrounds with the \nucleus{} shielding. Each histogram shows the rate of events with deposited energy between 0 and 1\,keV in the \cawo{} array of target detectors. Uncertainties are statistical only. The impact of the \nucleus{} setup passive materials is shown on the right panel. The left panel illustrates the complementary rejection brought by the use of the IV and COV. The \cenns{} RoI is indicated by the light gray filled area.}
    \label{fig:Gamma-induced_shielding_mitigation}
\end{figure*}

This confirms the need for adding a gamma ray shield to suppress this contribution down to <\,100\,\rate{}.
The optimal amount of Pb material in the external and internal gamma shields results from a trade-off between (i) the production of secondary particles when exposed to cosmic ray-induced neutrons and muons (see section~\ref{subsubsec:CR_mitigation}) and (ii) the total reduction of ambient gamma-ray backgrounds when combined with the COV.
A thickness of 5\,cm of Pb followed from extensive \Geant{} simulation studies.
It gives a factor $\sim$\,50 reduction of the event rate in the \cawo{} target detectors, while the rest of the \nucleus{} setup passive materials (see figure~\ref{fig:nucleus_layout}d and figure~\ref{fig:nucleus_layout}e) makes an additional factor $\sim$\,10 reduction.
The right panel of figure~\ref{fig:Gamma-induced_shielding_mitigation} illustrates the effect of the \nucleus{} IV and COV detectors.
It particularly points out the essential role of the COV in reducing the environmental gamma ray contribution down to sufficiently low levels.
As opposed to the atmospheric neutron backgrounds, varying the COV energy threshold was found to have little impact on the ambient gamma ray rejection performances as long as it stays under the 100 $\mathrm{keV_{ee}}$ range.
This is mostly because in this energy regime, gamma rays rarely interact through Compton scattering but rather deposit their energy through photo-electric absorption.
The IV is a very thin detector (see figure~\ref{fig:nucleus_layout}f), which makes it insensitive to the interactions of environmental gamma rays. As expected, it does not improve the rejection of this background component.

%
%
%%%%%%%%%%%%%%%%%%%%%%%%%%
%%% MAT. RADIOPURITY %%%%%
%%%%%%%%%%%%%%%%%%%%%%%%%%
%
\subsubsection{Mitigation of material radioactivity}\label{subsubsec:material_radioactovity_mitigation}
The work reported in this section deals with modeling the backgrounds sourcing from the radioactive contamination measured in the main \nucleus{} setup components (see section~\ref{subsec:material_radiopurity}) and how the veto detectors help mitigating them.
The generation of the primary particles follows a different approach to that used for the previously discussed external sources of background.
It was done in a similar fashion than for the modeling of the ambient gamma-ray background in the VNS (see section~\ref{subsec:gamma_at_vns}), using a custom event generator which homogeneously distributes the position of the decaying primordial radionuclides in each of the \nucleus{} setup volumes as implemented in the simulation framework geometry (see figure~\ref{fig:nucleus_layout}). 
The decay chain of each of the considered primordial radionuclides ($^{232}$Th, $^{235}$U, $^{238}$U and $^{40}$K) was simulated in one step, unless the activity of a sub-chain was measured (see table~\ref{tab:Screening_result}). 
In the latter case, the simulation of the full decay chain was broken down into the corresponding sub-chains until reaching the first stable nucleus. 
The modeling of the material radioactive contamination contribution to the expected backgrounds in the \nucleus{} cryogenic target detectors involved the analysis of $\sim$\,100 different simulation datasets, each having a primary event statistics corresponding to at least a 500-day exposure time.
After analysis, the rate and energy spectrum of surviving events associated to each of these datasets were computed using the corresponding chain or sub-chain activities as reported in table~\ref{tab:Screening_result}.
In the following results, chain or sub-chain measurements resulting in an upper limit were disregarded.
The impact of this approximation was evaluated by assuming instead such upper limits as effective activity values, giving a modest 5\% increase in the expected total background rate.

The left panel of figure~\ref{fig:material_radioactivity_mitigation} shows the expected residual rate of events in the \cawo{} target detectors between 0 and 1\,keV, by combining the contributions of each screened component and each measured chain or sub-chain. 
It also describes how the use of the IV and COV detectors can mitigate this background component.
Unsurprisingly, the COV is the most efficient veto detector. 
It reduces the background rate by an order of magnitude down to $\sim$\,10\,\rate{} levels, once again demonstrating its essential role in the \nucleus{} whole background mitigation strategy. 
The combination of all anti-coincidence selection criteria, including the IV, brings the total background rate to below 10\,\rate{} in the \cenns{} RoI. 
The right panel of figure~\ref{fig:material_radioactivity_mitigation} quantifies the contribution of each of the \nucleus{} setup components, still showing the impact of the different veto detectors. 
For some of them (e.g. the detector clamps), the application of all veto rejection criteria left no surviving events, which resulted in a null contribution to the predicted backgrounds.
The \bfourc{} neutron cryogenic shield contributes most, as it is (i) among the most radioactive of all screened materials and (ii) located close to the cryogenic detectors. 
The second largest contribution comes from the intrinsic radioactivity of the target detectors, for which none of the veto detectors shows a visible effect because they are located outside this specific background source.
This last observation particularly points out the necessity of having a more careful radiopurity selection for all materials used in the building of the cryogenic target detectors than for the rest of the experiment.
The fact that the expected \nucleus{} backgrounds from material radioactivity are here limited by the intrinsic radiopurity of the \cawo{} target detector material demonstrates a good management of material screening and selection in the design of the \nucleus{} experiment.
\begin{figure}[ht!]
    \centering
    \includegraphics[width=0.49\textwidth]{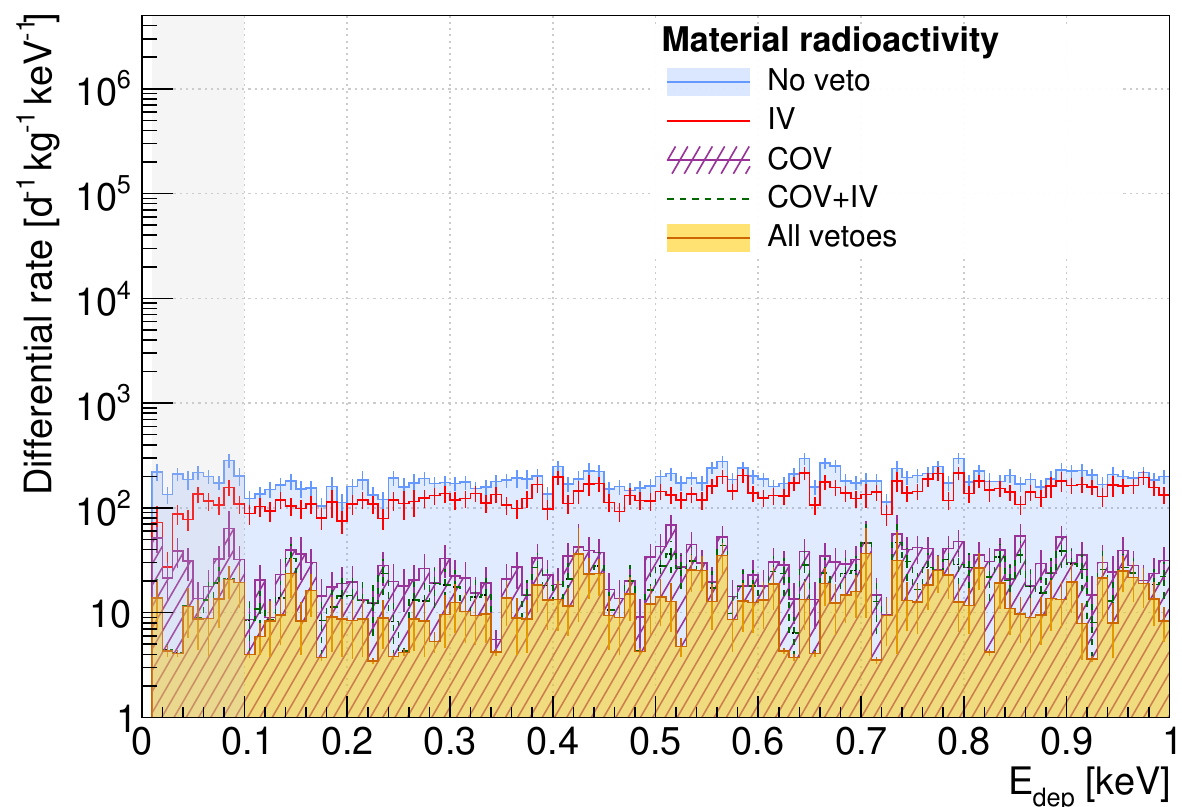}
    \hfill
    \includegraphics[width=0.49\textwidth]{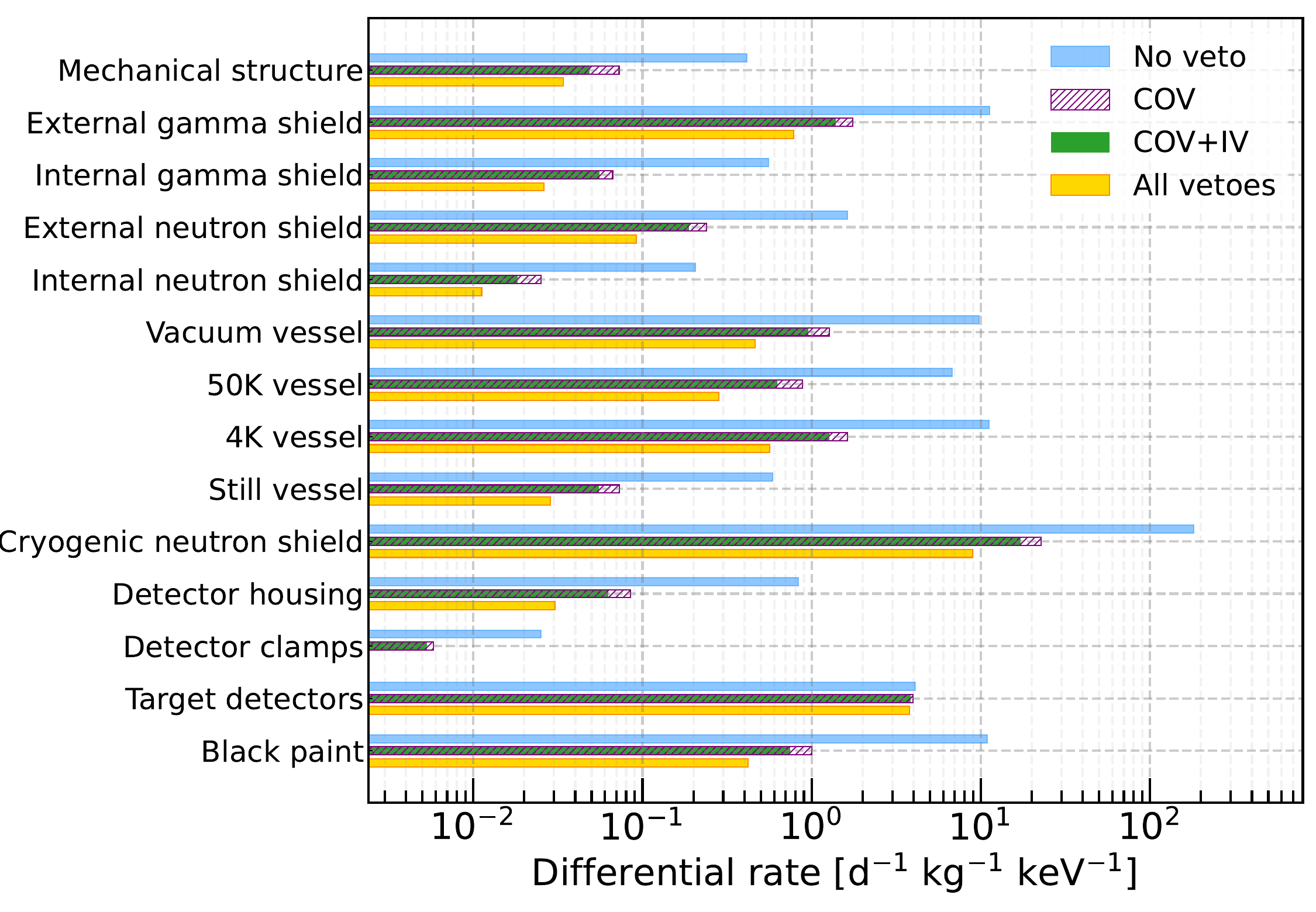}
    \caption{Modeling of the material radioactivity contribution to the \nucleus{} expected backgrounds. The top panel shows how the veto detectors can help mitigating this background component. The \cenns{} RoI is indicated by the light gray filled area. Uncertainties are statistical only. The bottom panel breaks down the contribution of each of the screened \nucleus{} components. Each histogram shows the rate of events with deposited energy between 0 and 1\,keV in the \cawo{} array of target detectors.}
    \label{fig:material_radioactivity_mitigation}
\end{figure}

%%% ================================================ Background budget ==================================================
\subsection{Residual background rates and spectra at the VNS} \label{subsec:bck_budget}
The prediction of the residual backgrounds in the \nucleus{} cryogenic target detectors at the VNS must take into account the building attenuation to the cosmic ray-induced neutron and muon components. 
By design, the modeling of the ambient gamma ray and material radioactivity contributions at the VNS remains unchanged with respect to the results presented in the previous section.
The prediction of the atmospheric neutron-induced backgrounds started from the open-air simulation dataset used in section~\ref{subsubsec:CR_mitigation}. 
Monte Carlo events passing the analysis selection criteria were reweighted such that the energy distribution of their associated primary neutrons matched the neutron ambiance spectrum as predicted in the VNS instead of the surface \textit{Gordon} spectrum (see left panel of figure~\ref{fig:nFluxVNSroom}). 
Changes in the angular distribution of neutrons or the production of secondary particles due to the building were therefore ignored. 
As opposed to the atmospheric neutrons, the production of secondary particles (neutrons and gamma rays) from the interaction of atmospheric muons in the building materials can make a significant difference in the prediction of the event rates at low energies with respect to e.g. simply rescaling an open-air modeling with the measured attenuation at the VNS.
A detailed simulation of muon transport through the building up to the target detectors is then required.
Because of computing time considerations, the atmospheric muon component was modeled using a two-step simulation.
In a first step, the atmospheric muon custom primary generator was used to transport muons through the VNS building only, scoring the information of any particle tracks reaching the VNS room.
A second simulation step made use of a bootstrapping method, where the distribution of scored particle tracks was resampled and transported several times through the \nucleus{} shielding up to the target detectors.
This allowed to obtain a decent statistical accuracy with a much lower computational cost than using a brute force analog simulation.
The achieved Monte Carlo primary event statistics was for each of the simulated background components larger than at least a year of exposure time. 
The analysis of the simulation datasets applied all possible veto criteria, which selects \cenns{}-like events as described in section~\ref{subsubsec:bck_prediction_methods}. 
Given the very low event rates, Monte Carlo statistical uncertainties anyhow had to be taken into account when assessing the uncertainty associated to the background prediction presented here. 
For > 10 counts statistics, the mean event rate was assumed to be Poisson distributed. 
In the case of smaller statistics, upper limits at 90\% confidence level were estimated using the Feldman-Cousins prescription for the computation of confidence intervals~\cite{feldmancousins1998}.
\begin{table*}[ht!]
     \caption{Expected background budget in the \nucleus{} target detectors at the VNS after implementation of veto requirements and all selection criteria. The event rates are expressed in milli count per day (mcpd). The quoted uncertainties are statistical only. The corresponding systematic uncertainties can be directly calculated from the flux relative uncertainties quoted in table~\ref{tab:Flux_Uncertainties}. The \cenns{} expected count rates are also given for comparison. They were estimated for a 100\% operating cycle for the B1 and B2 reactors, as well as a more realistic 80\% operating cycle, which takes into account scheduled reactor stops for maintenance and refueling.
    }
    \label{tab:bckg_budget}
    
    \makebox[\textwidth][c]{\parbox{1.3\textwidth}{
    \addtolength{\leftskip} {-2cm}
    \addtolength{\rightskip}{-2cm}
    \renewcommand{\arraystretch}{1.2}
    \centering
    \small
    \begin{NiceTabular}{@{}@{\extracolsep{\fill}} p{0.25\textwidth}ccccc @{}}
    \toprule
    &\multicolumn{2}{c}{\textbf{CaWO$_4$}} && \multicolumn{2}{c}{\textbf{Al$_2$O$_3$}}\\
    \textbf{Background source} & \textbf{10-100\,eV} & \textbf{0.1-1\,keV} && \textbf{10-100\,eV} & \textbf{0.1-1\,keV} \\
    \cmidrule{2-3} \cmidrule{5-6} 
    Atm. muons & <\,14 &  <\,18 && <\,11 &  <\,23 \\  
    Atm. neutrons & ${131\pm6}$ & ${305\pm8}$ && ${48\pm4}$ & ${168\pm6}$ \\
    Env. gamma rays & ${6\pm3}$ & ${60\pm11}$ && <\,3 & ${66\pm11}$ \\
    Material radioactivity & ${7.1\pm1.1}$ & ${80\pm5}$ && ${4.3\pm1.1}$  & ${86\pm6}$  \\
    \textbf{Total} & $\mathbf{144 \pm 6}$ & $\mathbf{445 \pm 14}$ && $\mathbf{52 \pm 4}$ & $\mathbf{320 \pm 14}$ \\
    \bottomrule
    \\
    \textbf{CE$\nu$NS signal} & 218.2 & 60.7 && 8.4 & -\\
    \end{NiceTabular}
    }}
\end{table*}

Table~\ref{tab:bckg_budget} details the expected \nucleus{} background budget both in the \cawo{} and \alo{} arrays of target detectors. 
Background event rates are reported in the \cenns{} RoI between 10 and 100\,eV as well as higher energies between 0.1 and 1\,keV. 
Unsurprisingly, the cosmic ray-induced atmospheric neutron component is predicted to be the main contributor. 
This observation is a direct consequence of the VNS very shallow overburden, which does not completely cut off the neutron component of the surface cosmic ray-induced radiations.
The unavoidable need for a high-Z material gamma-ray shield, which especially leads to the production of secondary neutrons, complicates the mitigation of this background source.
Another limitation comes from the compactness of the \nucleus{} shielding, which must fit the dimensions of the VNS. 
Still, section~\ref{subsec:shield_bck_reduction} demonstrated that a clever combination of passive materials and active detectors allows to achieve a very good suppression of all the particle background sources expected at the VNS.
In particular, a signal-to-background ratio greater than or equal to one can reasonably be expected in the \cawo{} array of target detectors.
Additionally, thanks to the compact size of the muon veto, a reasonable rate of 325\,cps is expected in the muon veto inducing a 2 to 8\,\% dead time depending on the detector response time \cite{NUCLEUSlbr2025}. The COV is expected to observe a rate of the order of 10\,cps, leading to a negligible impact on the dead time. 
\added{The proportion of deferred events by more than \SI{100}{\micro\second} (for example decaying particles and nuclei, mentioned in section \ref{sec:nucleus_geant4}) remains negligible and represents only 1\,\% of the surviving background.}
At the exception of the atmospheric neutrons, the \cawo{} and \alo{} detectors do not show significant differences in their expected event rates with respect to the achieved statistical uncertainties.
This is because the \nucleus{} shielding design strongly rejects background events induced by atmospheric muons, environmental gamma rays and material radioactivity, making it difficult to reach high Monte Carlo statistics using analog simulations.
However, merging the ambient gamma rays and material radioactivity contributions  (see e.g. table~\ref{tab:bckg_budget}) shows that the gamma-induced rates in the \cawo{} and \alo{} detectors are very similar in the sub-keV recoil energy range.
Atmospheric neutron-induced event rates are predicted to be larger in the \cawo{} target detectors than in the \alo{} detectors. 
The reason for this is twofold: (i) the \cawo{} material exhibits a larger neutron interaction cross-section and (ii) the elastic scattering process, which dominates the total neutron interaction cross-section, kinematically produces much smaller recoils in \cawo{} than in \alo{}.

The top panel of figure~\ref{fig:residual_bck_spectra} shows how the background sources at the VNS contribute to the expected residual event spectrum between 0 and 10\,keV in the \cawo{} detectors.
They all result in an overall flat contribution to the spectrum, except the atmospheric neutron component, which imprints a sharp rise at low energies.
This rise originates from elastic scattering of 0.001-1\,MeV neutrons off tungsten nuclei, which survived the passive shields and veto detectors, and made their way up to the target detectors.
Interestingly, the residual background spectrum also features fluorescence peaks in the 2-3\,keV region, which are predicted to follow from the ionization of M shell electrons in tungsten atoms.
As further shown by the top panel of figure~\ref{fig:residual_bck_spectra}, these fluorescence peaks are mostly induced by atmospheric neutrons and muons.
If confirmed, they could be an interesting tool to monitor and complement the target detector energy calibration in the\,keV energy regime and below~\cite{abele2025}.

\begin{figure}[ht!]
\centering
    \includegraphics[width=1.\linewidth]{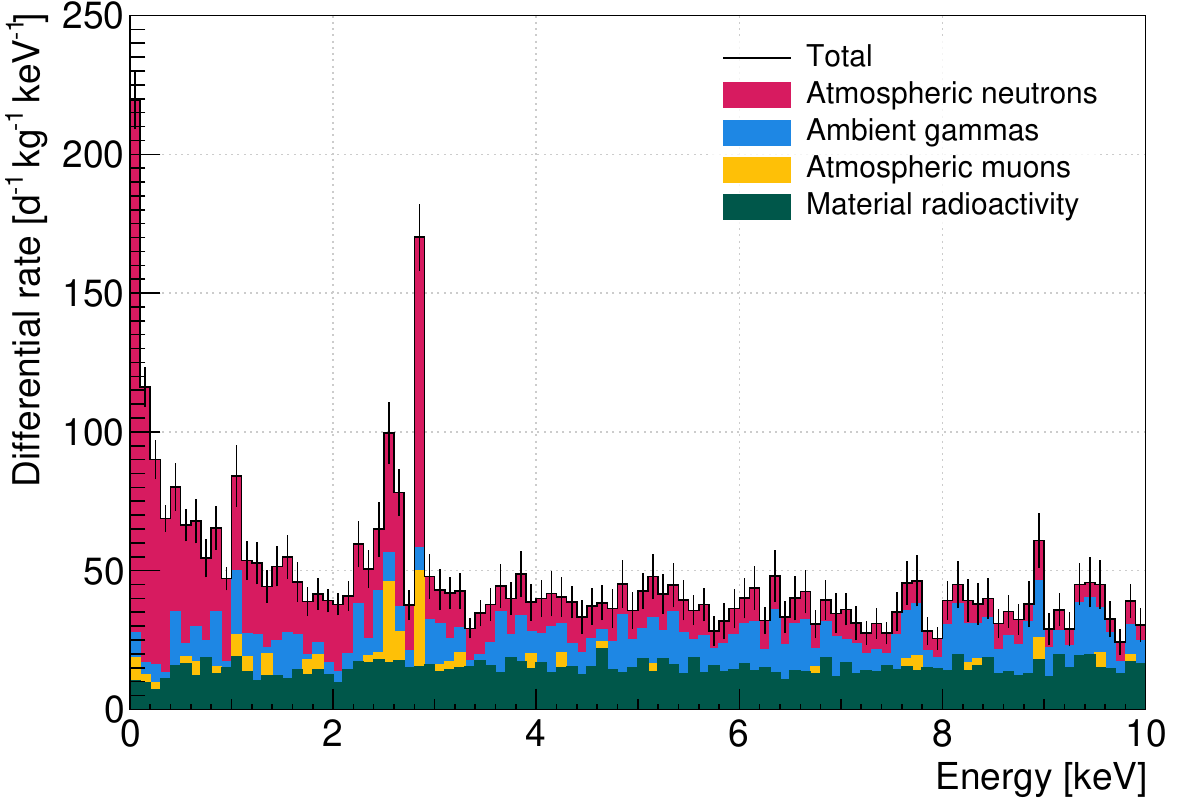}
    \includegraphics[width=1.\linewidth]{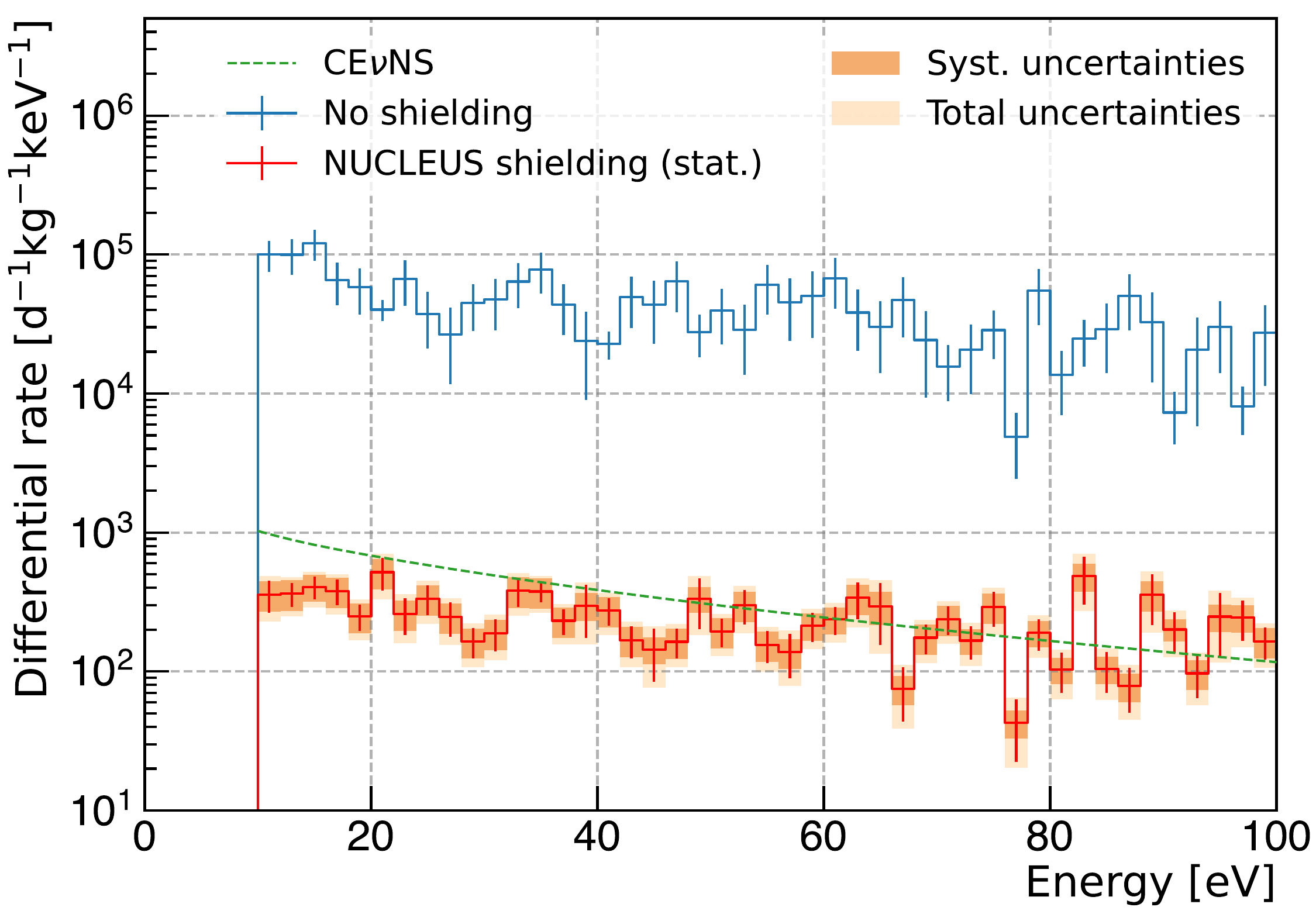}
    \caption{Residual background spectra predicted in the \nucleus{} array of \cawo{} detectors at the VNS. The top panel shows the contribution of the different sources of particle backgrounds to the total spectrum between 0 and 10\,keV. The bottom panel illustrates the \nucleus{} shielding total rejection power between 10 and 100\,eV. The residual spectrum (solid red line) is compared to the expected \cenns{} differential recoil spectrum computed assuming a 80\% average operating cycle for both the B1 and B2 reactors (green dashed line). The impact of the combined systematic uncertainties sourcing from the normalization of the different background components (dark orange boxes) on the total estimated uncertainties (light orange boxes) is also shown.\\
    }
    \label{fig:residual_bck_spectra}
\end{figure}
The bottom panel of figure~\ref{fig:residual_bck_spectra} shows how the background events distribute within the \cenns{} RoI between 10 and 100\,eV. The systematic uncertainties shown in the bottom panel of figure~\ref{fig:residual_bck_spectra} and in figure~\ref{fig:S_N_COVThr} are propagated using a conservative \text{min/max} method. The background contributions from all sources are scaled up (respectively down) by their uncertainties and summed to obtain the upper (respectively lower) bound of the plotted uncertainty band.
Combined to the VNS overburden, a $\mathrm{10^2-10^3}$ particle background rejection power is expected from the \nucleus{} shielding, bringing the residual rate of events in the targeted \rate{100} range and at the same level than the expected \cenns{} \added{on \cawo{}} signal \added{(largely dominated by \cenns{} on W)}. 
As already discussed in section~\ref{subsubsec:CR_mitigation}, the rejection of atmospheric neutron-induced backgrounds is sensitive to the applied energy threshold for defining a detector hit in the COV.
All the results presented in table~\ref{tab:bckg_budget} and figure~\ref{fig:residual_bck_spectra} were obtained using a 1 $\mathrm{keV_{ee}}$ energy threshold. 
Although already demonstrated in the past~\cite{armengaud2010,armengaud2012}, achieving such thresholds with the six COV HPGe detectors in a dry cryostat however remains a challenge.
Figure~\ref{fig:S_N_COVThr} therefore shows the impact of varying the COV energy threshold on the expected signal-to-background ratio in the \cenns{} RoI. 
An energy threshold in the $\cal{O}$(10 $\mathrm{keV_{ee}}$) range, as demonstrated during a first commissioning of the experiment~\cite{NUCLEUSlbr2025}, degrades the expected signal-to-background ratio, which nevertheless remain acceptable.
Lack of precise knowledge about the absolute neutron flux at the VNS, which limits the precision of the present background prediction, is illustrated by the gray band on figure~\ref{fig:S_N_COVThr}.
For a 1 $\mathrm{keV_{ee}}$ COV energy threshold, it expands the signal-to-background ratio to the [0.9-1.5] range at the 68\% confidence level.
Finally, demonstrating a 10\,eV energy threshold in all the \cawo{} and \alo{} target detectors in the early stage of the experiment is challenging~\cite{NUCLEUSlbr2025}. 
Figure~\ref{fig:S_N_COVThr} also shows how the expected signal-to-background ratio would deteriorate when a 20\,eV energy threshold is assumed.
\begin{figure}[ht!]
\centering
    \includegraphics[width=1.\linewidth]{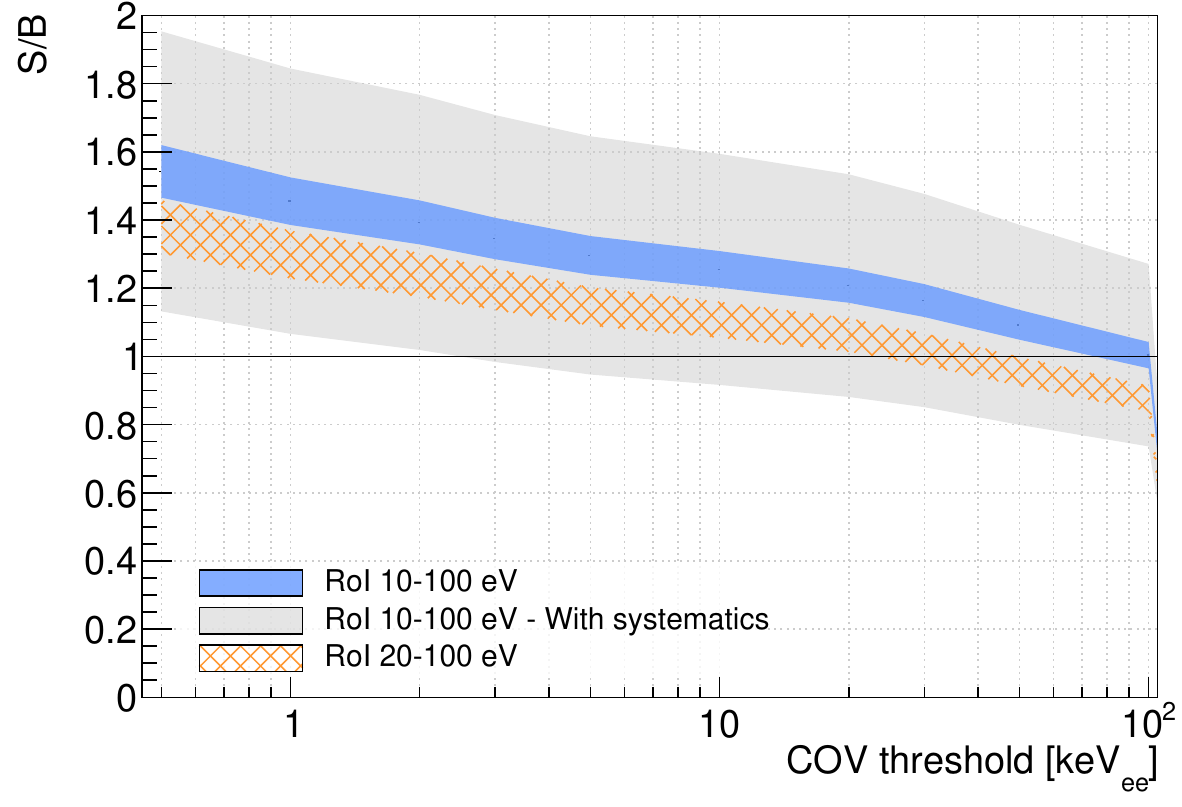}
    \caption{Expected signal-to-background ratio bands as a function of the applied COV energy threshold. Two possible RoIs are also reported for illustrating the impact of the cryogenic target detector energy threshold. The gray band includes on top of the Monte Carlo statistical uncertainties the systematic uncertainties for normalizing each background component. See text for further details.
    }
    \label{fig:S_N_COVThr}
\end{figure}

%%% ===================================================================================================================================
%%% ================================================ Conclusions ==================================================
%%% ===================================================================================================================================

\section{Conclusions}\label{sec:conclusion}
This article comprehensively detailed the particle background mitigation strategy of the \nucleus{} experiment for the detection of coherent elastic neutrino-nucleus scattering at the Chooz nuclear power plant. 
The experimental signature of \cenns{}, which is a simple nuclear recoil with energies reaching at most $\cal{O}$(1\,keV) for reactor antineutrinos, together with the compactness and the very low $\sim$\,\SI{3}{\m\we} overburden offered by the VNS experimental site pose serious challenges for the mitigation of particle backgrounds, especially those being induced by cosmic rays. 
A thorough characterization of the background radiation environment of the VNS was used to (i) fine-tune and validate the \nucleus{} \Geant{} simulation framework for the modeling and transport of external background sources at the experimental site and (ii) estimate the residual background rates and spectra in the \nucleus{} cryogenic target detectors. 
The \nucleus{} shielding system, which combines several layers of passive shields together with a muon veto and an arrangement of cryogenic vetoes installed in close proximity to the target detectors, is predicted to suppress by more than two orders of magnitude particle-induced backgrounds, giving a \cenns{} signal-to-background ratio $\gtrsim$\,1 in the 10-100\,eV region of interest. 
A key element of the \nucleus{} background mitigation strategy is the high purity germanium COV, which efficiently rejects gamma ray-related backgrounds, thereby minimizing the necessary amount of Pb material and allowing the experimental setup to fit within the small footprint of the VNS.
The COV was also found to nicely complement the muon veto in rejecting muon-induced backgrounds, and even offers additional rejection to neutrons. 
Additionally to external radiations, background sources internal to the experimental apparatus were also investigated in details. 
The material radioactivity contribution to the background budget was kept to a minimal thanks to a careful selection of the most important setup materials, with a very modest impact comparable to the external ambient gamma rays.
The total background budget was found to be dominated by a residual cosmic ray-induced fast neutron contribution.
Although the \bfourc{} cryogenic shield was found to be very effective in suppressing neutron-related events in the \cenns{} region of interest, the shallow VNS overburden makes it difficult to fully suppress this background component.
Furthermore, the suppression of the cosmic ray-induced neutron backgrounds was found to significantly depend on the COV energy threshold.
Unsurprisingly, a threshold as low as \SI{1}{\keVee} is highly desirable in order to take full advantage of the COV in reducing these backgrounds.

Although already very detailed and comprehensive, this version of the \nucleus{} particle background model is not yet ready to provide a precise prediction of the residual background rate and energy spectrum for the \nucleus{} \cenns{} physics run, as it still features some limitations and physics data to be fine-tuned on. 
Among the most important ones are the absolute normalization of the neutron ambiance at the VNS, for which a measurement is currently missing, and the full completion of the material radioactivity contribution, which lacks the screening measurement of a few elements in very close proximity to the target detectors. 
The data to be collected during the technical run at Chooz, scheduled for early next year, will be then essential for fine-tuning the model in these aspects. 
The reliability of the \Geant{} standard physics lists for the transport and interaction of particles down to the sub-keV energy regime is another major question mark regarding the robustness of the present modeling. 
The different commissioning data recently acquired at TU Munich are a very valuable source of information to test and benchmark the \nucleus{} \Geant{} background model down to the\,keV energy range.
However, the overwhelming LEE background component still prevents such a study in the $\cal{O}$(100\,eV) region.
The forthcoming technical run at Chooz, which will test and challenge the LEE background mitigation strategy of the \nucleus{} cryogenic detector design, will hopefully provide valuable information to validate the particle background model down to the \cenns{} region of interest.
Finally, the brute force analog simulations performed throughout this work required tens, even hundreds, millions of hours of computing time. 
The strong attenuation at play in the transport of particles, together with the scarcity of energy depositions in the sub-keV energy regime, are a computational challenge for achieving a high enough Monte Carlo statistical precision.
They will certainly require the use of new statistical methods, such as variance reductions techniques, to be not limited by the accuracy and the precision of the particle background model in the future extraction of a \cenns{} signal at Chooz.

\section*{Acknowledgments}
This work has been financed by the CEA, the INFN, the ÖAW and partially supported by the TU Munich and the MPI für Physik. NUCLEUS members acknowledge additional funding by the DFG through the SFB1258 and the Excellence Cluster ORIGINS, by the European Commission through the ERC-StG2018-804228 “NU-CLEUS”, by the P2IO LabEx (ANR-10-LABX-0038) in the framework ”Investissements d’Avenir” (ANR-11-IDEX-0003-01) managed by the Agence Nationale de la Recherche (ANR), France, by the Austrian Science Fund (FWF) through the ”P34778-N, ELOISE" (\href{https://www.fwf.ac.at/en/research-radar/10.55776/P34778}{DOI:10.55776/P34778}), and by the Max-Planck-Institut für Kernphysik (MPIK), Germany.
The NUCLEUS collaboration thanks the CRESST collaboration for providing input data on the \cawo{} detector radiopurity, as well as the LNGS staff running the \textsc{Stella} facility for the material screening campaigns.


\bibliography{sn-bibliography}% common bib file

\end{document}